\documentclass[12pt]{article}
\usepackage{lineno}
\usepackage{epsfig}
\usepackage{amsmath}
\usepackage{hhline}
\usepackage{amssymb}
\usepackage{times}
\usepackage{cite}
\usepackage{array}
\usepackage{multirow}

\usepackage{color}
\definecolor{rltred}{rgb}{0.75,0,0}
\definecolor{rltgreen}{rgb}{0,0.5,0}
\definecolor{rltblue}{rgb}{0,0,0.75}

\newlength{\dinwidth}
\newlength{\dinmargin}
\begin{document}  

%\linenumbers
%\renewcommand\linenumberfont{\normalfont\small\bfseries}
\setlength\linenumbersep{1cm}
\resetlinenumber

% Some useful tex commands
\newcommand{\pom}{{I\!\!P}}
\def\lsim{\,\lower.25ex\hbox{$\scriptstyle\sim$}\kern-1.30ex%
\raise 0.55ex\hbox{$\scriptstyle <$}\,}
\newcommand{\qsq}{\ensuremath{Q^2} }
\def\submJournal#1{{#1}}
\def\Journal#1#2#3#4{{#1} {\bf #2} (#3) #4}
\def\EJC{{\em Eur. Phys. J.} {\bf C}}
\def\NIM{\em Nucl. Instrum. Methods}
\def\CPC{\em Comp. Phys. Commun.}
\def\EJ{{\em Eur. Phys. J.}}
\def\NP{{\em Nucl. Phys.}}
\def\PL{{\em Phys. Lett. }}
\def\PR{{\em Phys. Rev. }}
\def\PRL{\em Phys. Rev. Lett.}
\def\ZP{{\em Z. Phys.}}
\def\RMP{{\em Rev. Mod. Phys. }}
\def\JHEP{{\em J. High Energy Phys.} {\bf JHEP}}
\def\SJNP{{\em Sov. J. Nucl. Phys.}}
\def\SPJ{{\em Sov. Phys. JETP}}
\newcommand{\etal}{{\em et al.}}

\newcommand{\arxiv}[1]{%
%\ifpdf
%\href{http://arXiv.org/ps/#1}{\mbox{[#1]}}
%\else
\mbox{[#1]}}
%\fi}

\newcommand{\arxivnobrackets}[1]{%
%\ifpdf
%\href{http://arXiv.org/ps/#1}{\mbox{#1}}
%\else
\mbox{#1}}
%\fi}

\newcommand\figuresname{Figures}
\newcommand\tablesname{Tables}
\renewcommand{\eqref}[1]{Eq.~(\ref{#1})}
\newcommand{\figref}[1]{\figurename~\ref{#1}}
\newcommand{\figsref}[2]{\figuresname~\ref{#1} and~\ref{#2}}
\newcommand{\figrange}[2]{\figuresname~\ref{#1}--\ref{#2}}
\newcommand{\tabref}[1]{\tablename~\ref{#1}}
\newcommand{\tabrange}[2]{\tablesname~\ref{#1}--\ref{#2}}
\newcommand{\secref}[1]{Section~\ref{#1}}
\newcommand{\secrange}[2]{Sections~\ref{#1}--\ref{#2}}
\fboxsep5pt
\fboxrule0.4pt
\newcommand{\Pom}{{I\hspace{-0.75ex} P}}
\newcommand{\etamax}{%
\ensuremath{%
{\eta_{\rm max}}}}
\newcommand{\xpom}{%
\ensuremath{%
{x_\pom}}}
\newcommand{\xpomeron}{%
\ensuremath{%
{x_\pom}}}
\newcommand{\logxpomeron}{%
\ensuremath{%
{\log_{10}(x_\pom)}}}
\newcommand{\ftd}{%
\ensuremath{%
{F_2^D}}}
\newcommand{\pthat}{%
\ensuremath{%
{\hat{p}_T}}}
\newcommand{\zpom}{%
\ensuremath{%
{z_\pom}}}
\newcommand{\zpomeron}{%
\ensuremath{%
{z_\pom}}}
\newcommand{\zpomeronjets}{%
\ensuremath{%
{z_\pom^{\rm jets}}}}
\newcommand{\xgam}{%
\ensuremath{%
{x_\gamma}}}
\newcommand{\xgamma}{%
\ensuremath{%
{x_\gamma}}}
\newcommand{\xgammajets}{%
\ensuremath{%
{x_\gamma^{\rm jets}}}}
\newcommand{\xgammajetspar}{%
\ensuremath{%
{x_{\gamma, {\rm PL}}^{\rm jets}}}}
\newcommand{\ptjet}{%
\ensuremath{%
{E_T^{\rm jet1}}}}
\newcommand{\etjetone}{%
\ensuremath{%
{E_T^{\rm *,jet1}}}}
\newcommand{\ptjetone}{%
\ensuremath{%
{E_T^{\rm *,jet1}}}}
\newcommand{\ppbar}{%
\ensuremath{%
{p\bar{p}}}}
\newcommand{\etjettwo}{%
\ensuremath{%
{E_T^{\rm *,jet2}}}}
\newcommand{\deltaeta}{%
\ensuremath{%
{\left| \Delta\eta_{\rm jet}\right|}}}
\newcommand{\deltaetastar}{%
\ensuremath{%
{\left| \Delta\eta^*_{\rm jet}\right|}}}
\newcommand{\etajetlab}{%
\ensuremath{%
{\eta_{\rm jet}^{\rm lab}}}}
\newcommand{\meanetajetlab}{%
\ensuremath{%
{\langle \eta_{\rm jet}^{\rm lab}\rangle}}}
\newcommand{\meanetajet}{%
\ensuremath{%
{\langle \eta_{\rm jet}\rangle}}}
\newcommand{\mjj}{%
\ensuremath{%
{M_{12}}}}

\begin{titlepage}

\noindent
DESY 07-018 \\
February 2007

\vspace{2cm}

\begin{center}
\begin{Large}

{\bf Tests of QCD Factorisation in the Diffractive Production of Dijets in Deep-Inelastic Scattering and Photoproduction
     at HERA}

\vspace{2cm}

H1 Collaboration

\end{Large}
\end{center}

\vspace{2cm}

\begin{abstract}

\noindent
Measurements are presented of differential dijet cross sections in diffractive
photoproduction ($Q^2<0.01$~GeV$^2$) and deep-inelastic scattering processes
(DIS, $4<Q^2<80$~GeV$^2$). The event topology is given by $ep \rightarrow e X
Y$, in which the system~$X$, containing at least two jets, is separated from a
leading low-mass proton remnant system~$Y$ by a large rapidity gap.   The dijet
cross sections are compared with NLO QCD predictions based on diffractive
parton densities previously obtained from a QCD analysis of inclusive
diffractive DIS cross sections by H1.  In DIS, the dijet data are well
described, supporting the validity of QCD factorisation.  The diffractive DIS
dijet data are more sensitive to the diffractive gluon density at high
fractional parton momentum than the measurements of inclusive diffractive DIS.
In photoproduction, the predicted dijet cross section has to be multiplied by a
factor of approximately $0.5$ for both direct and resolved photon interactions to
describe the measurements.  The ratio of measured dijet cross section to NLO
prediction in photoproduction is a factor $0.5 \pm 0.1$ smaller than the same
ratio in DIS.  This suppression is the first clear observation of QCD hard
scattering factorisation breaking at HERA.  The measurements are also compared
to the two soft colour neutralisation models SCI and GAL.  The SCI model
describes diffractive dijet production in DIS but not in photoproduction. The
GAL model fails in both kinematic regions.
\end{abstract}

\vspace{1.5cm}

\begin{center}
Submitted to \EJC
\end{center}

\end{titlepage}

\begin{flushleft}
%-- H1AUTS Author list by names 
%-- Status: Tue Nov 14 09:30:57 CET 2006  Number of authors = 289 

A.~Aktas$^{10}$,               %DESY-LEFT      09/06           Aktas               
V.~Andreev$^{24}$,             %LPI -PD        8/88            Andreev             
T.~Anthonis$^{4}$,             %ANTW-ST        11/99           Anthonis            
B.~Antunovic$^{25}$,           %MPIM-ST        09/03           Antunovic           
S.~Aplin$^{10}$,               %DESY-PD        01/04           Aplin               
A.~Asmone$^{32}$,              %ROME-ST        07/2            Asmone              
A.~Astvatsatourov$^{4}$,       %BRUX-PD        07/04           Astvatsatourov      
A.~Babaev$^{23, \dagger}$,     %ITEP-LEFT      12/05           Babaev              
S.~Backovic$^{29}$,            %PODG-PD        03/2            Backovic            
A.~Baghdasaryan$^{37}$
% ...

%-- H1 Institutes 
\bigskip{\it
% ...
$ ^{2}$ Vinca  Institute of Nuclear Sciences, Belgrade, Serbia \\
% ...
$ ^{4}$ Inter-University Institute for High Energies ULB-VUB, Brussels;
          Universiteit Antwerpen, Antwerpen; Belgium$^{ c}$ \\
% ...
$ ^{10}$ DESY, Hamburg, Germany \\
% ...
$ ^{23}$ Institute for Theoretical and Experimental Physics,
          Moscow, Russia$^{ k}$ \\
 $ ^{24}$ Lebedev Physical Institute, Moscow, Russia$^{ e}$ \\
 $ ^{25}$ Max-Planck-Institut f\"ur Physik, M\"unchen, Germany \\
% ...
$ ^{29}$ Faculty of Science, University of Montenegro,
          Podgorica, Montenegro$^{ e}$ \\
% ...
$ ^{32}$ Dipartimento di Fisica Universit\`a di Roma Tre
          and INFN Roma~3, Roma, Italy \\
% ...
$ ^{37}$ Yerevan Physics Institute, Yerevan, Armenia \\
% ...
$ ^{\dagger}$ Deceased \\

\bigskip
% ...
$ ^c$ Supported by FNRS-FWO-Vlaanderen, IISN-IIKW and IWT
      and  by Interuniversity
Attraction Poles Programme,
      Belgian Science Policy \\
% ...
$ ^e$ Supported by the Deutsche Forschungsgemeinschaft \\
% ...
$ ^k$ Partially Supported by Russian Foundation
      for Basic Research,  grants  03-02-17291
      and  04-02-16445 \\
% ...
}

\end{flushleft}

\newpage

\section{Introduction}
It can be shown in Quantum Chromodynamics (QCD)  that the cross section for
diffractive processes in deep-inelastic $ep$ scattering (DIS) factorises into
universal diffractive parton density functions (DPDFs) of the proton and
process-dependent hard scattering cross sections (QCD
factorisation)~\cite{Collins}.  Diffractive parton densities have been
determined from QCD fits to inclusive diffractive cross section measurements in
DIS by H1~\cite{h1f2d94,h1f2d97}. It was found that most of the momentum of the
diffractive exchange is carried by gluons.

Final state configurations for which a partonic cross section is perturbatively
calculable include dijet and heavy quark production, which are directly
sensitive to the diffractive gluon distribution. Previous measurements of
diffractive dijet production in DIS~\cite{h1oldjets,disjets} have been found to
be described by leading order (LO) Monte Carlo (MC) QCD calculations based on
the factorisation approach that use the diffractive parton densities
from~\cite{h1f2d94} and include parton showers to simulate higher order
effects.  However, using the same diffractive parton densities in LO QCD
calculations overestimates the cross section for single-diffractive dijet
production in \ppbar{} collisions at the Tevatron by approximately one order of
magnitude~\cite{tevjets}.  This discrepancy has been attributed to the presence
of the additional beam hadron remnant in \ppbar{} collisions, which leads to
secondary interactions.  The suppression, often characterised by a `rapidity
gap survival probability,' cannot be calculated perturbatively but has been
parameterised in various ways (see,
e.g.,~\cite{bjorkengap,telaviv1,telaviv,cox,kaidalov}).

An alternative approach to diffractive scattering is taken by soft colour
neutralisation models in which diffraction is described by partonic hard
scattering processes with subsequent reconfiguration of colour between the
final state partons.  One of these models is the Soft Colour Interaction
model~\cite{sci} which, when tuned to describe inclusive diffractive HERA
measurements, also gives a reasonable description~\cite{scitev} of diffractive
Tevatron data~\cite{tevjets,scidata1,scidata2,scidata3,scidata4,scidata5}.

The transition from deep-inelastic scattering to hadron-hadron scattering can
be studied at HERA by comparing scattering processes in DIS and in
photoproduction.  In photoproduction, the beam lepton emits a quasi-real photon
which interacts with the proton ($\gamma p$ collision).  Processes in which the
photon participates directly in the hard scattering are expected to be similar
to the deep-inelastic scattering of highly virtual photons (`point-like
photon'). In contrast, processes in which the photon is first resolved into
partons which then engage in the hard scattering resemble hadron-hadron
scattering.  These resolved photon processes can produce gluon-gluon and
gluon-quark final states, which are present in \ppbar{} collisions but
negligible in DIS.  Furthermore, they have an additional hadronic remnant which
opens up the possibility of remnant-remnant interactions.  QCD factorisation is
proven for diffractive DIS, is also expected to hold for direct photon
interactions in diffractive photoproduction~\cite{Collins}, but not for
resolved processes.  Previous comparisons of diffractive photoproduction dijet
data with LO MC models showed consistency with QCD factorisation within large
uncertainties~\cite{h1oldjets}.

Measurements of diffractive D$^*$ meson (charm) production are well described
by next-to-leading order (NLO) QCD calculations and by LO Monte Carlo models
based on diffractive parton densities in both DIS~\cite{dstar, dstarzeus,
zeusnlo, h1roger} and photoproduction~\cite{h1roger}. However, these
measurements suffer from large statistical uncertainties of the data.

In this paper, a more precise test of QCD factorisation for diffractive dijet
production in DIS and photoproduction is presented.  Measurements of
diffractive dijet cross sections are compared with NLO QCD predictions based on
recently published diffractive parton densities~\cite{h1f2d97} from H1.  In
addition, the dijet cross sections are also compared with two versions of the
LO soft colour interaction model.  The data were collected with the H1 detector
at HERA in the years 1996 and 1997.  For photoproduction the integrated luminosity is
increased by one order of magnitude with respect to previous results. For DIS,
the same data sample is used as in a previous measurement~\cite{disjets}.  Jets
are defined using the inclusive $k_T$ cluster algorithm~\cite{kt} with
asymmetric cuts on the jet transverse energies to facilitate comparisons with
NLO predictions~\cite{poetter,kkasym}.  Apart from the different ranges for the
photon virtuality, the DIS and photoproduction measurements are performed in
the same kinematic range to allow the closest possible comparison of the
results.

\section{Kinematics}
\begin{figure}[htb]
\begin{center}
\includegraphics[width=0.5\linewidth]{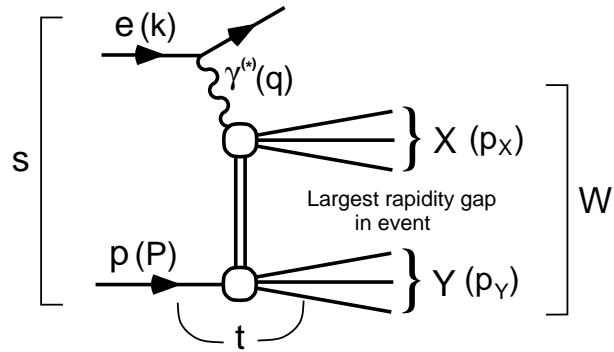}
\caption{Illustration of the generic diffractive process $ep \rightarrow e X
Y$. The systems $X$ and $Y$ are separated by the largest gap in the rapidity
distribution of the final state hadrons.}
\label{fig_generic}
\end{center}
\end{figure}

The generic diffractive positron-proton interaction $ep \rightarrow e X Y$ is
illustrated in \figref{fig_generic}. The positron (4-momentum $k$) exchanges a
photon ($q$) which interacts with the proton ($P$).  The produced final state
hadrons are, by definition, divided into the systems $X$ and $Y$, separated by
the largest gap in the hadron rapidity distribution relative to the
$\gamma^{(*)}p$ collision axis in the photon-proton centre-of-mass frame. The
system $Y$ lies in the outgoing proton beam direction.

\begin{figure}[htb]
\begin{center}
\textbf{a)}
\includegraphics[width=0.4\linewidth]{desy07-018_fig2a.eps}
\hspace{0.4 cm}
\textbf{b)}
\includegraphics[width=0.5\linewidth]{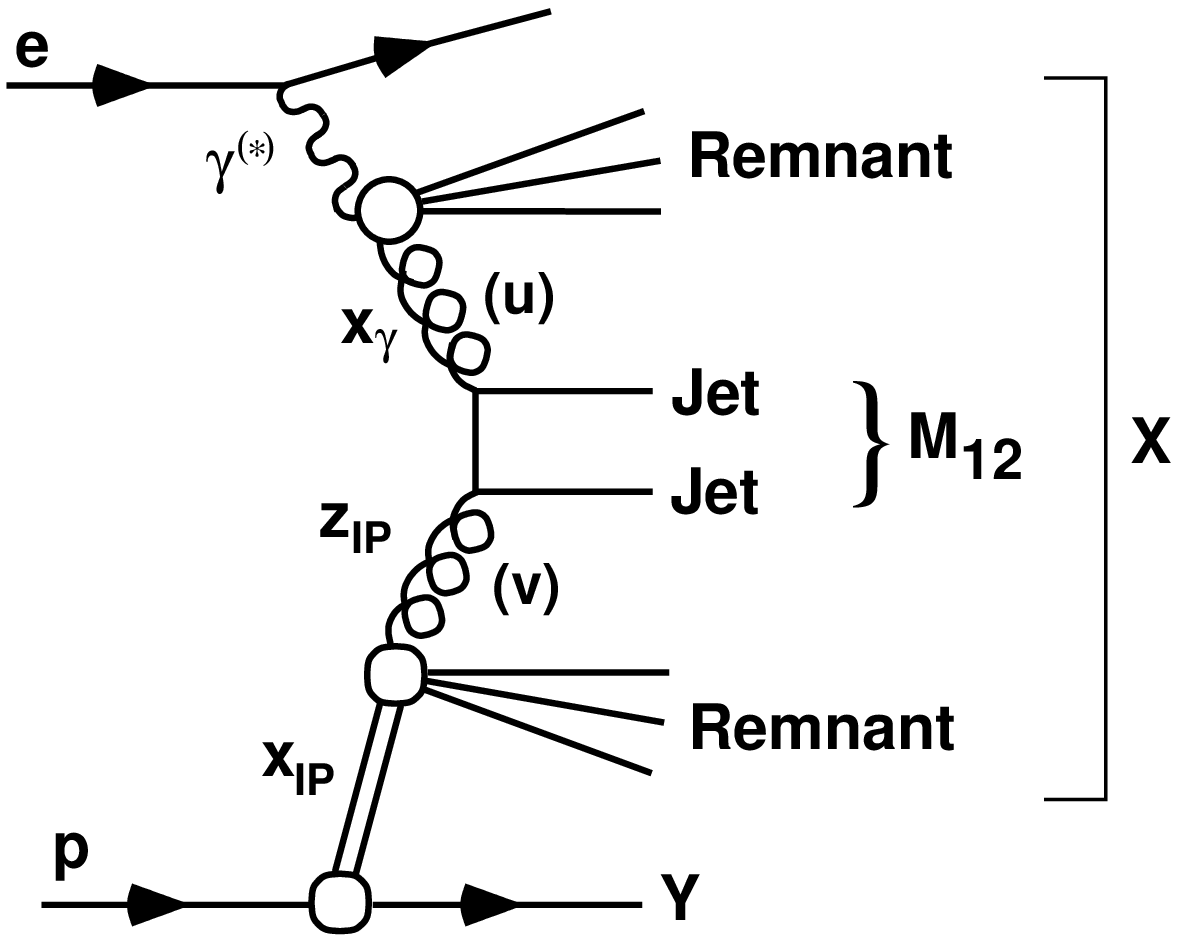} \\
\caption{Leading order diagrams for diffractive dijet production at HERA with
the diffractive exchange depicted as a pomeron (\Pom).  a) Direct (point-like)
photon process (photon-gluon fusion), b) resolved (hadron-like) photon
process.}
\label{fig_feynBGF}
\end{center}
\end{figure}

Examples of direct and resolved photon processes with dijets in the final state
are depicted in \figref{fig_feynBGF}.  Resolved processes give a large
contribution in photoproduction but are suppressed in DIS. The diffractive
exchange in these diagrams is depicted as a pomeron (\Pom).

The usual DIS kinematic variables are defined as:
\begin{equation}
Q^2 \equiv -q^2\ , \qquad y\equiv
\frac{q\cdot P}{k\cdot P}\ , \qquad x \equiv \frac{Q^2}{2 P\cdot q}\ .
\end{equation}
These three variables are related through $Q^2 \approx sxy$, in which $s \equiv
(k+P)^2$ is the fixed $ep$ centre-of-mass energy squared.  The invariant mass
of the photon-proton system $W$ is given by
\begin{equation}
W = \sqrt{(q+P)^2} \approx \sqrt{y\,s - Q^2}\ .  \label{eq_W}
\end{equation}
With $p_X$ and $p_Y$ representing the 4-momenta of the systems $X$ and $Y$, we
define
\begin{equation}
M^2_X\equiv p_X^2\ , \quad M^2_Y\equiv p_Y^2\ , \quad t\equiv (P-p_Y)^2\ ,\quad 
\xpom \equiv \frac{q\cdot (P-p_Y)}{q\cdot P}\ .
\end{equation}
The quantities $M_X$ and $M_Y$ are the invariant masses of the systems $X$ and
$Y$, $t$ is the squared four-momentum transferred at the proton vertex and
\xpom{} represents the fraction of the proton beam momentum transferred to the
system $X$. Diffractive events are characterised by small values of \xpom{}
($\lsim 0.05$). With $u$ and $v$ denoting the four-momenta of the two partons
(\figref{fig_feynBGF}b) or photon and parton (\figref{fig_feynBGF}a) entering
the hard subprocess,  the dijet system has squared invariant mass
\begin{equation}
M^2_{12} = (u+v)^2\ .
\end{equation}
The fractional longitudinal momenta carried by the partons from the photon
(\xgamma) and the diffractive exchange (\zpomeron) are given by
\begin{equation}
\xgamma=\frac{P\cdot u}{P\cdot q}\ , \qquad \zpom=\frac{q\cdot v}{q\cdot
(P-p_Y)}\ .
\end{equation}
The measurements are performed in the region $\xpom<0.03$, $-t<1$~GeV$^2$ and
\mbox{$M_Y<1.6$~GeV}, where the cross section is dominated by scattering
processes in which the proton stays intact.

\section{Diffractive Dijet Production in the Factorisation Approach}
\label{sec:models}
In the QCD factorisation approach, diffractive $ep$ dijet cross sections are
calculated according to the formula
\begin{eqnarray}
\lefteqn{{\rm d} \sigma(ep \rightarrow e + 2~{\rm jets} + X' + Y) \ =
\sum_{i,j} \int\!\!{\rm d}y \  f_{\gamma/e}(y)
 \int\!{\rm d}\xgam \, f_{j/\gamma}(\xgam, \mu_{F}^2) \ \  \times}
 \nonumber \\
& &{} \times\, 
\int\!\!{\rm d}t 
\int\!\!{\rm d}\xpom 
\int\!\!{\rm d}\zpom \  
d\hat{\sigma}(ij \rightarrow 2~{\rm jets}) \
f_i^D(\zpom,\mu_{F}^2,x_\pom,t),
\label{equ:diffdijetfull}
\end{eqnarray}
in which the sum runs over all contributing partons, $f_{\gamma/e}$ is the
photon flux from the positron and $f_{j/\gamma}$ are the photon parton
densities. For direct photon interactions, $f_{j/\gamma} = \delta(1-\xgamma)$.
The partonic cross sections are denoted by $\hat{\sigma}$ and $f_i^D$ are the
diffractive parton densities of the proton.  The factorisation scale $\mu_F$ is
assumed to be identical at the photon and proton vertices.  In the present
analysis, the jet transverse energy is larger than $Q$ for most of the data and
is therefore used as the factorisation scale and as the renormalisation scale
both in DIS and in photoproduction.  The variable $X'$ denotes the part of the
hadronic system $X$ which is not contained in the two jets. 
 
\label{sec:h1dpdf}
The H1 Collaboration has determined diffractive parton densities from QCD fits
to inclusive diffractive DIS data in~\cite{h1f2d94,h1f2d97}.  In the
parameterisations used for these fits, the \xpom{} and $t$ dependences of the
diffractive parton distributions were factorised from the dependences on the
scale $\mu_F$ and the fractional parton momentum $\zpom$:
\begin{equation}
f_i^D(\zpom,\mu_F^2,x_\pom,t) = f_{\pom}(x_\pom,t) \ \ f_{i,\pom}(\zpom,\mu_F^2).
\label{reggefac}
\end{equation}
The factor $f_{\pom}(x_\pom,t)$ was parameterised as suggested by Regge theory.
The dependence on $\zpom$ was parameterised at a starting scale and evolved to
the scale at which the inclusive data were measured using the DGLAP evolution
equations~\cite{dglap,dglapnlo}.  The inclusive diffractive DIS
data~\cite{h1f2d94,h1f2d97} are well described using this approach. For
$\xpom>0.01$, small additional contributions from sub-leading meson (`reggeon')
exchange have to be taken into account to describe the data.

The H1 Collaboration has published QCD fits to two different data sets of
inclusive diffractive DIS events. In a first analysis~\cite{h1f2d94}, data
taken in the year 1994 were used to extract the LO `H1 fit 2' parton densities
which have been used previously in comparisons with diffractive dijet
production in DIS at HERA and at the Tevatron.  A second analysis was based on
the larger data samples of the years 1997--2000~\cite{h1f2d97}.  The fit
in~\cite{h1f2d97} led to the NLO `H1~2006~Fit~A' and NLO `H1~2006~Fit~B' DPDFs
which both give a good description of inclusive diffraction, and which are the
basis of  the dijet predictions in this paper.  The two sets of parton
densities differ mainly in the gluon density at high fractional parton momentum,
which is poorly constrained by the inclusive diffractive scattering data. The
gluon density of Fit A is peaked at the starting scale at high fractional
momentum and that of Fit B is flat.

\section{Next-to-leading Order QCD Calculations}
\label{sec_nlo}
Existing programs which calculate NLO QCD partonic cross sections for dijet
production in inclusive DIS and photoproduction can be adapted to calculate
cross sections in diffraction. For DIS, the DISENT \cite{disent} program is
used, as suggested in \cite{hautmann}.  It was demonstrated in
\cite{poetter,nlocomp,nlojet} that dijet calculations using this program agree
very well with the results from other
programs~\cite{disaster,jetvip,mepjet,nlojet}.  The program by Frixione
et~al.~\cite{frixione} is used for photoproduction. 

The two NLO programs are adapted to calculate diffractive cross sections
according to the following procedure.  The cross section at fixed $\xpom$ and
$t=0$ is calculated by reducing the nominal proton beam energy by a factor
$\xpom$. Since the $x_\pom$ and $t$ dependences of the DPDFs are assumed to
factorise from the $\zpom$ and $\mu_F$ dependences, the proton PDFs can be
replaced by the parton densities of the diffractive exchange
$f_{i,\pom}(\zpom,\mu_F^2)$. The cross sections are multiplied by
$f_{\pom}(x_\pom,t)$, integrated between $t=-1$~GeV$^2$ and the maximum
kinematically allowed value of $t$.  In the same way, a $\approx 3\%$
contribution from Reggeon exchange is calculated.  Kinematic effects on the
partonic configurations arising from finite values of $t$ are neglected.  To
compare the results with the measured cross sections in the region
$\xpom<0.03$, the results are integrated over $\xpom$.

The diffractive dijet cross sections of the modified programs have been
compared at the LO tree level with predictions of the Monte Carlo generator
RAPGAP~\cite{RAPGAP} (see also \secref{sec_rapgap}).  Good agreement has been
found for both DIS and photoproduction, indicating that the diffractive
extension works correctly.  The diffractive NLO predictions agree with
independent calculations in both DIS and photoproduction~\cite{kkdis,kkgp}.

For the NLO predictions in this paper, the recent H1 2006 DPDFs are used and
the 2-loop strong coupling $\alpha_s(M_Z)$ is set to $0.118$; the same value is
used in the evolution of the parton densities~\cite{h1f2d97}. The
renormalisation scale is set to the transverse energy of the leading parton jet
in the photon-proton centre-of-mass frame.  In DISENT it is not possible to
change the factorisation scale on an event-by-event basis. It is therefore set
to the average $E_T$ of the leading jet observed in the DIS measurement
($6.2$~GeV). Variations of the QCD renormalisation scale by factors $0.5$ and $2$
in DISENT result in changes of the predicted dijet cross section by
approximately $+24\%$ and $-17\%$, respectively, integrated over the DIS
kinematic range specified in \tabref{tab_xsdef}.  Varying the factorisation
scale by factors $0.5$ and $2$ leads to changes of the predicted dijet cross
section by approximately $+8\%$ and $-7\%$, respectively.  In the Frixione
program for photoproduction, the factorisation and renormalisation scales are
fixed to be equal.  Variations of the scales by factors $0.5$ and $2$ change the
predicted cross section by approximately $+33\%$ and $-21\%$, respectively,
integrated over the photoproduction kinematic range specified in
\tabref{tab_xsdef}.  In photoproduction, the GRV HO photon PDFs~\cite{GRVgamma}
are used. Photon parton densities are not used in DISENT.

The calculated NLO parton jet cross sections are corrected for the effects of
hadronisation.  The corrections, defined as
\begin{equation}
\left( 1+\delta_{\rm had} \right)_i = \left( \frac{\sigma^{\rm hadron}_{\rm
    dijet}}{\sigma^{\rm parton}_{\rm dijet}} \right)_{\!\!\!i},
\end{equation}
are determined for both DIS and photoproduction in every measurement bin $i$
using the two Monte Carlo generators RAPGAP with Lund string fragmentation and
HERWIG~\cite{herwig} with cluster fragmentation.  The HERWIG program was
extended to diffraction in the manner described above for the NLO programs and
uses LO diffractive parton densities.  For the parton level cross section
$\sigma^{\rm parton}_{\rm dijet}$ the jet algorithm operates on the final state
partons after the parton shower cascade.  The hadronisation correction is
calculated as the mean of the corrections obtained from RAPGAP and HERWIG. The
difference between the two corrections serves as an error estimate.  In DIS,
the hadron level dijet cross section does not differ significantly from the
cross section at the parton level.  In photoproduction, the hadron level cross
section is lower than the parton level cross section by $10\%$ on
average. The correction is particularly large at high $\xgamma$ where
contributions with $\xgamma \approx 1$ at the parton level are smeared towards
lower values due to hadronisation. The estimated uncertainty on $(1+\delta_{\rm
had})$ is $20\%$ for $\zpomeronjets>0.8$ in DIS and less than $10\%$ in all
other measurement bins.  It is listed in \tabrange{tab_xsdis1}{tab_xsgp2}.

The uncertainty on the parton densities arising from experimental and
theoretical uncertainties in the fit to inclusive diffractive data are much
smaller than the QCD scale uncertainties of the dijet predictions and are
neglected.  The NLO corrections increase the LO cross section by factors
$1.9$ and $1.7$ on average in DIS and photoproduction,
respectively.  This large correction is due to the low transverse energy of the
jets. 

\section{Soft Colour Neutralisation}
An approach conceptually different from that of diffractive parton densities is
provided by soft colour neutralisation models. In these models, diffractive
scattering is described by DIS or photoproduction hard scattering processes
with subsequent colour rearrangements between the final state partons.  This
soft reconfiguration leaves the parton momenta unchanged and can produce colour
singlet systems which are separated by a large rapidity gap. 

The Soft Colour Interaction model (SCI)~\cite{sci} contains one free parameter,
the colour rearrangement probability, which was fitted to \ftd{} measurements.
A refined version of the model (GAL)~\cite{gal} uses a generalised area law for
the colour rearrangement probability.  Both versions of the model give a
reasonably good description~\cite{scitev} of HERA inclusive diffractive cross
sections and of diffractive processes at the
Tevatron~\cite{tevjets,scidata1,scidata2,scidata3,scidata4,scidata5}. 

Predictions for diffractive dijet production in the SCI and GAL models are
obtained using the LO generator programs LEPTO~\cite{leptosci} and
PYTHIA~\cite{PYTHIA} for the DIS and photoproduction kinematic regions,
respectively. Higher order QCD effects are simulated using  parton showers.
The calculations are based on the CTEQ5L LO parton densities of the
proton~\cite{cteq5l}.

\section{Experimental Procedure}
\subsection{H1 detector}
A detailed description of the H1 detector can be found in~\cite{H1det}.  Here,
a brief account of the components most relevant to the present analysis is
given. The H1 coordinate system convention defines the outgoing proton beam
direction as the positive $z$ axis, also referred to as the `forward'
direction.  The polar angle $\theta$ is measured relative to this axis and the
pseudorapidity is defined as $\eta \equiv -\ln \tan (\theta/2)$.

The central $ep$ interaction region is surrounded by two large concentric jet
drift chambers, two $z$ chambers, and two multi-wire proportional chambers
(MWPCs), located inside a $1.15$~T solenoidal magnetic field.  Charged particle
momenta are measured by the drift chambers in the range \mbox{$-1.5<\eta<1.5$}
with a resolution of \mbox{$\sigma(p_T)/p_T \simeq 0.005\,
  p_T/$GeV~$\oplus~0.015$.} The
\mbox{MWPCs} provide fast trigger information based on the signals of charged
particles.  In the central and forward region the track detectors are
surrounded by a finely segmented Liquid Argon calorimeter (LAr). It consists of
an electromagnetic section with lead absorbers and a hadronic section with
steel absorbers and covers the range \mbox{$-1.5 <\eta <3.4$.} The energy
resolution is $\sigma(E)/E \simeq 0.11/\sqrt{E/{\rm GeV}}$ for electromagnetic
showers and $\sigma(E)/E \simeq 0.50/\sqrt{E/{\rm GeV}}$ for hadrons, as
measured in test beams.  The backward region $-4 < \eta < -1.4$ is covered by a
lead/scintillating fibre calorimeter (SPACAL)~\cite{spacal} consisting of an
electromagnetic and a hadronic section.  The electromagnetic part is used to
identify and measure the scattered positron in DIS events with an energy
resolution of $\sigma(E)/E \simeq 0.07/\sqrt{E/{\rm GeV}}~\oplus~0.01$.  In front of the
SPACAL, the Backward Drift Chamber (BDC) provides track segments of charged
particles with a resolution of $\sigma(r)=0.4$~mm and $r \sigma(\phi)=0.8$~mm.

The forward region is instrumented with the Forward Muon Detector (FMD) and the
Proton Remnant Tagger (PRT).  Three double layers of drift chambers of the FMD
are used to detect particles with pseudorapidities in the range $1.9 < \eta <
3.7$.  The FMD can also detect particles from larger pseudorapidities which
reach the detector after undergoing secondary scattering with the beam-pipe.
The PRT consists of a set of scintillators surrounding the beam pipe at
$z=26$~m and covers the region $6 < \eta < 7.5$.

The $ep$ luminosity is measured with a precision of $1.5\%$ via the Bethe-Heitler
Bremsstrahlung process $ep \rightarrow ep \gamma$, the photon being detected in
a crystal calorimeter at $z=-103$ m.  A further crystal calorimeter at
$z=-33$~m is used as a small angle positron detector to measure the scattered
positron in photoproduction events. 

\subsection{Event selection}
The data correspond to an integrated luminosity of $18$~pb$^{-1}$ and
were taken in the 1996 and 1997 running periods,
in which HERA collided $820$~GeV protons with $27.5$~GeV positrons. The
measurements are described in detail in~\cite{sesphd}.  

The photoproduction data are collected using a trigger which requires the
scattered positron to be measured in the small angle positron detector, at
least three tracks to be reconstructed in the central jet chambers and an event
vertex to be identified.  A veto cut requiring less than $0.5$~GeV of energy
deposited in the photon detector of the luminosity system suppresses initial
state radiation and coincidences with Bremsstrahlung events. The geometrical
acceptance of the small scattering angle positron detector limits the photon
virtuality to $Q^2<0.01$~GeV$^2$ and the photon-proton centre-of-mass energy to
$165<W<242$~GeV.

DIS events are collected using a trigger which requires the scattered positron
to be detected in the backward electromagnetic calorimeter (SPACAL), an event
vertex to be identified and at least one high transverse momentum track ($p_T >
0.8$~GeV) to be measured in the central jet chambers. Several cuts are applied
on the SPACAL positron candidate to reduce background from photons and hadrons.
The electromagnetic cluster energy is required to be larger than $8$~GeV and
requirements are imposed on the width of the electromagnetic shower, the
containment in the electromagnetic section of the SPACAL and an associated
track segment in the BDC. DIS events with initial state QED radiation are
suppressed by requiring the summed $E-p_z$ of all final state particles
including the positron to be greater than $35$~GeV.  The range in the photon
virtuality is restricted to $4<Q^2<80$~GeV$^2$.  The photon-proton
centre-of-mass energy $W$ is restricted to the same range as for
photoproduction.

Diffractive events are selected in the same way as for the inclusive
diffractive cross section measurement~\cite{h1f2d97} used for the extraction of
the DPDFs.  No signals above noise thresholds are allowed in the FMD or PRT.
In the LAr, no cluster with an energy of more than $400$~MeV is allowed in the
region $\eta > 3.2$. These selection criteria ensure that the gap between the
systems $X$ and $Y$ spans at least the region $3.2 < \eta < 7.5$, and restrict
$M_Y$ and $t$ to approximately $M_Y < 1.6$~GeV and \mbox{$-t < 1$~GeV$^2$.} A
cut $\xpom < 0.03$ further reduces non-diffractive contributions. 

The hadronic system $X$ is measured in the LAr and SPACAL calorimeters and the
central tracking system. Calorimeter cluster energies and track momenta are
combined into hadronic objects using an algorithm which avoids double
counting~\cite{combobj}.  Jets are formed from the hadronic objects, using the
inclusive $k_T$ cluster algorithm~\cite{kt} with a distance parameter of unity
in the photon-proton rest frame, which is identical to the laboratory frame for
photoproduction up to a Lorentz boost along the beam axis.  The $p_T$
recombination scheme is used, which leads to massless jets.  At least two jets
are required, with transverse energies $\etjetone > 5$~GeV and $\etjettwo >
4$~GeV for the leading and sub-leading jet, respectively.\footnote{The `*'
denotes variables in the photon-proton rest frame.} The jet axes of the two
leading jets are required to lie within the region $-1 < \etajetlab < 2$, well
within the acceptance of the LAr calorimeter.  The final selection yields $1365$
events in photoproduction and $322$ events in DIS.

\subsection{Kinematic reconstruction}
\label{sec_rec}

\subsubsection{Reconstruction of DIS events}
In the DIS analysis, the energy $E_e$ and the polar angle $\theta_e$ of the
scattered positron are measured in the backward calorimeter SPACAL and $y$ and
$Q^2$ are reconstructed according to
\begin{equation}
y = 1- \frac{E_e}{E^0_e}\, \sin^2\frac{\theta_e}{2}\ , \qquad Q^2=4
E^0_e E_e\cos^2\frac{\theta_e}{2}\ ,
\end{equation}
in which $E^0_e$ is the positron beam energy.  The invariant mass $M_X$ of the
hadronic system $X$ is reconstructed from the energies $E_i$ and the momenta
$\vec{p_i}$ of all hadronic objects:
\begin{equation}
M_X^2 = (\sum_{i \in X} E_i)^2 - ( \sum_{i \in X} \vec{p}_i)^2\ . \label{eq_mx}
\end{equation}
The photon-proton centre-of-mass energy $W$ is reconstructed according to
\eqref{eq_W} and the variable \xpom{} is given by 
\begin{equation}
     \xpom = \frac {Q^2 + M_X^2 }{Q^2 + W^2}\ .
\end{equation}
The estimators \xgammajets{} and \zpomeronjets{} of the fractional momenta of
the partons entering the hard sub-process are reconstructed as
\begin{equation}
\xgammajets = \frac{\sum_{i=1}^2\, \left( E_{{\rm jet}\,i}^*-p^*_{z, {\rm jet}\,i}
\right)}{\sum_{i \in X }\, \left( E_i^*-p^*_{z,i} \right)}\ , \qquad
\zpomeronjets = \frac{ \qsq + M^2_{12}}{\qsq + M^2_X}\ .
\end{equation}

\subsubsection{Reconstruction of photoproduction events}
In the photoproduction analysis, the energy $E_e$ of the scattered positron is
measured in the small scattering angle positron detector and $y$ is
reconstructed according to
\begin{equation}
y= 1-E_e/E^0_e. 
\end{equation}
The estimators \xgammajets{} and \zpomeronjets{} are reconstructed as 
\begin{equation}
\xgammajets = \frac{\sum_{i=1}^2\, \left( E_{{\rm jet}\,i}-p_{z, {\rm jet}\,i}
\right)}{2\,y\,E^0_e}, \qquad
\zpomeronjets = \frac{\sum_{i=1}^2\, \left( E_{{\rm jet}\,i}+p_{z,
    {\rm jet}\,i}
\right)}{2\,\xpom\,E_p},
\end{equation}
in which $E_p$ is the incident proton beam energy.  The variable \xpom{} is
reconstructed according to 
\begin{equation}
     \xpom = \frac{\sum_{i \in X} \left(E_i+p_{z,i}\right)}{2\, E_p}.
\end{equation}
The reconstruction of $\zpomeronjets$ and $\xpom$ is different from the DIS
case due to the large contribution of resolved photon processes.

\subsection{Monte Carlo simulations}
\label{sec_rapgap}
Monte Carlo programs are used in the analysis to correct the measured
distributions for detector effects.  The H1 detector response is simulated
using detailed detector simulation programs based on GEANT~\cite{geant3}.  The
Monte Carlo events are subjected to the same analysis chain as the data.

The main Monte Carlo generator used to correct the data distributions is
RAPGAP~\cite{RAPGAP}.  Events are generated according to a convolution of LO
diffractive parton densities with LO QCD matrix elements for the hard $2
\rightarrow 2$ subprocess.  The `H1 fit 2' DPDFs of~\cite{h1f2d94} are used.
RAPGAP includes resolved photon processes for which the partonic cross sections
are also convoluted with the parton densities of the photon.  In
photoproduction, the leading order GRV '94 parton distribution
functions~\cite{GRVgamma} are used, which were found to give a good description
of the effective photon structure function as measured by H1~\cite{Kaufmann}.
For DIS, processes with a resolved virtual photon are generated using the
SAS-2D parameterisation~\cite{sas2d}, which leads to a reasonable description
of inclusive dijet production~\cite{lowq2} in a similar $Q^2$ and $E_T$ range
to that studied here.  The PDFs are taken at the scale $\mu_F^2=\hat{p}_T^2 + 4
m_q^2$, where $\hat{p}_T$ is the transverse momentum of the emerging hard
partons and $m_q$ is the mass of the quarks produced. Higher order effects are
simulated using parton showers~\cite{PS} in the leading log($\mu$)
approximation.  The Lund string model~\cite{LUND} is used for hadronisation.
Photon radiation from the positron lines is simulated using the program
HERACLES~\cite{heracles}. The used RAPGAP version simulates only processes in
which the proton stays intact.

\subsection{Cross section measurement}
The data are first corrected for losses at the trigger level.  The trigger
efficiency is approximately $90\%$ in DIS, the losses being mainly due to the
tracking requirements.  In photoproduction the efficiency also depends on the
energy of the positron detected in the small scattering angle detector and
varies between $\approx 90\%$ at low $y$ and $\approx 50\%$ at high $y$.
Non-diffractive background migrating into the measurement region from
$M_Y>5~$GeV and large \xpom{} is statistically subtracted using inclusive dijet
production simulations (RAPGAP in DIS and PYTHIA in photoproduction).  The
subtracted background amounts to $3\%$ in photoproduction and $5\%$ in DIS.  Due to
the limited geometrical detector acceptance in the forward direction it is not
possible to distinguish an intact final state proton from one which dissociates
into a low-mass system $Y$.  Thus the measured cross section is defined to
include proton dissociation with $M_Y<1.6$~GeV. The correction factor for
migrations about the measurement boundary $M_Y=1.6$~GeV is determined using the
DIFFVM~\cite{DIFFVM} simulation of proton dissociation in the range
$m_p<M_Y<5$~GeV.  In the simulation, the ratio of elastic proton to proton
dissociation cross sections is assumed to be unity, in accordance with the
inclusive measurements of~\cite{h1f2d97,kapishin}.  The correction factors are
found to be $0.96\pm 0.04$ for the 1996 running period and $0.92\pm 0.05$ in
1997, the difference resulting from the degrading performance of the detectors
used to veto proton dissociation.  An additional factor $1.055\pm 0.014$ is
applied to account for the loss of diffractive events due to noise fluctuations
in the FMD.  This factor is determined using randomly triggered events.  A
correction of $5\%$ is applied to compensate for the removal of dijet
events in which a bremsstrahlung process is overlaid.  A small correction
($<1\%$) is applied to the measured DIS cross section to account for QED
radiation effects.

\begin{figure}[hhh]                    
\centering
\includegraphics[width=0.8\textwidth,keepaspectratio]{%
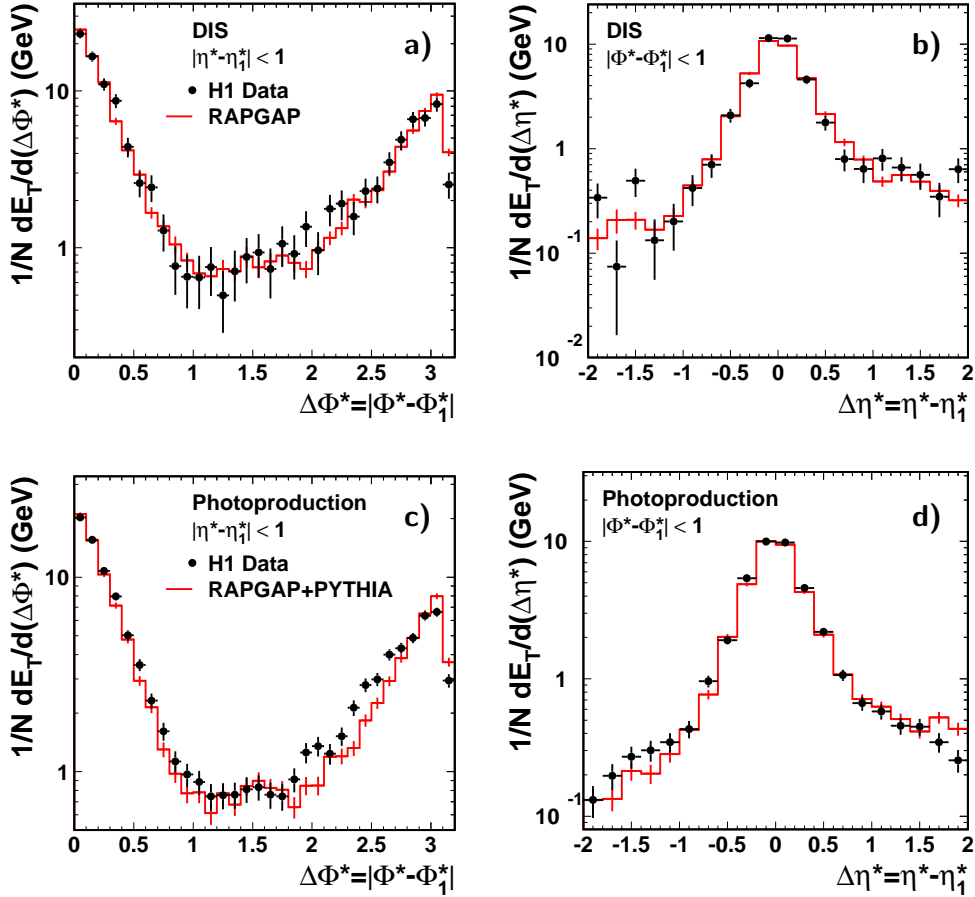} 
\caption{Average transverse energy flow per event around the leading jet axis
for diffractive dijets at the detector level in DIS (a and b) and
photoproduction (c and d).  The variables $\Delta \eta^*$ and $\Delta \Phi^*$
denote the distances from the axis of the leading jet in pseudorapidity and
azimuth in the photon-proton rest frame, respectively.  In a) and c) only
energy within one unit of pseudorapidity around the jet axis is included
whereas the profiles b) and d) include energy within one unit in azimuth around
the axis.}
\label{fig_profiles}
\end{figure}

The final jet cross sections are given at the hadron level.  The measured
distributions at the detector level are corrected for detector inefficiencies,
acceptances and migrations between measurement intervals in the reconstruction
using the RAPGAP Monte Carlo program and applying a bin-to-bin correction.
The simulation gives a good description of the shapes of all data distributions
and of the energy flow in the events.  \figref{fig_profiles} shows the
transverse energy flow around the axis of the leading jet for the selected
diffractive dijets in DIS (\figref{fig_profiles}a,b) and photoproduction
(\figref{fig_profiles}c,d).  A clear back-to-back structure is visible in the
$\Delta \Phi^*$ distribution.  The transverse energy flow in the jets as well
as in the region between the jets is reasonably well described by the
simulation.

According to the simulations, the detector level observables are well
correlated with the hadron level quantities.  Purities and
stabilities\footnote{`Purity' is defined as the fraction of Monte Carlo
simulated events reconstructed in a certain measurement interval that are also
generated in that bin. `Stability' is defined as the fraction of events
generated in a bin that are also reconstructed in that bin.} are larger than
$25\%$, the main source of migrations being the jet transverse energy
measurements.

The cross sections are measured in the kinematic region specified in
\tabref{tab_xsdef}.  The pseudorapidity range $-3<\eta^*<0$ in the
photon-proton rest frame used for the DIS measurement corresponds approximately
to the range $-1<\eta<2$ in the laboratory frame.

\setlength{\tabcolsep}{0.5cm}
\begin{table}                                                                  
\renewcommand{\arraystretch}{1.5}
\begin{center}
\begin{tabular}{|c|c|}
\hline
\bf Photoproduction & \bf DIS \\
\hline
\hline
$Q^2 < 0.01\ {\rm GeV}^2$ & $ 4 < Q^2 < 80 \ {\rm GeV}^2$ \\
\hline
\multicolumn{2}{|c|}{$ 165 < W < 242$ GeV} \\
\hline
\multicolumn{2}{|c|}{inclusive $k_T$ jet algorithm, distance
  parameter $=1$} \\
\hline
\multicolumn{2}{|c|}{$N_{\rm jet} \ge 2$} \\
\hline
\multicolumn{2}{|c|}{$\etjetone > 5$ GeV} \\
\hline
\multicolumn{2}{|c|}{$\etjettwo > 4$ GeV} \\
\hline
$-1 < \eta_{\rm jet(1,2)} < 2$ & $-3 < \eta_{\rm
jet(1,2)}^* < 0$ \\
\hline
\multicolumn{2}{|c|}{$\xpom <$ 0.03} \\
\hline
\multicolumn{2}{|c|}{$M_Y < 1.6$ GeV} \\
\hline
\multicolumn{2}{|c|}{$-t < 1$ GeV$^2$} \\
\hline
\end{tabular}
\end{center}
\caption{The kinematic ranges of the measured hadron level $ep$ cross sections.}
\label{tab_xsdef}
\end{table}

\subsection{Analysis of systematic uncertainties}
The following systematic errors on the measured cross sections arise from
experimental sources such as detector calibration uncertainties. The cross
section errors are estimated by repeating the analysis with variations in the
reconstruction of detector-simulated Monte Carlo events.
\begin{itemize}
\item A $4\%$ uncertainty in the absolute energy scale of the hadronic LAr
calorimeter in the jet $E_T$ range considered here~\cite{frankphd}  affects the
reconstruction of the hadronic final state.  The resulting uncertainty on the
measured cross section is $4\%$ in DIS and $8\%$ in photoproduction and is strongly
correlated between the data points.  The influence of this uncertainty in DIS
and in photoproduction is different due to the different reconstruction of
$\xpom$.  A $7\%$ uncertainty in the SPACAL hadronic energy scale affects the
cross sections by $1\%$.  The uncertainty in the fraction of the energy of the
reconstructed hadronic objects which is carried by tracks is $3\%$ and gives rise
to errors on the cross section of $4\%$ in photoproduction and $3\%$ in DIS, again
strongly correlated between data points.
\item 
The absolute SPACAL electromagnetic energy scale is known to $0.3\%$ for
scattered posi\-trons with $E_e=27.5$~GeV and $2.0\%$ at $E_e=8$~GeV.  The polar
scattering angle of the positron is measured to $1$~mrad precision.  The
uncertainties of the positron energy and angle measurements in DIS result in
cross section errors in the range of $4$ to $5\%$ for the energy uncertainty and
$2\%$ for the scattering angle.  In photoproduction, the uncertainty in the
knowledge of the acceptance and efficiency of the small angle positron detector
results in a cross section error of $5\%$ on average.
\item The uncertainties on the trigger efficiencies and the luminosity
measurement give rise to cross section uncertainties of $6\%$ and $1.5\%$,
respectively.
\item An uncertainty of $25\%$ in the fraction of events lost due to noise in the
FMD translates into a $1.3\%$ normalisation error on the cross section.
\end{itemize}

Systematic errors arising from uncertainties in the acceptance and migration
corrections are estimated by repeating the measurements with variations in the
kinematic dependences and other details of the Monte Carlo models within
experimentally allowed limits.

\begin{itemize}
\item The shapes of the following distributions in the RAPGAP simulation have
been varied: a) the $\zpomeron$ distribution in photoproduction has been
reweighted by factors $\zpomeron^{\pm 0.3}$ and \mbox{$(1-\zpomeron)^{\pm
0.3}$}; b) the $\ptjet$ distribution by $\hat{p}_T^{\pm 0.5}$ in both
photoproduction and DIS; c) the $\xpomeron$ distribution by $\xpomeron^{\pm
0.2}$ in photoproduction and $\xpomeron^{\pm 0.3}$ in DIS; d) the $\xgamma$
distribution by $\xgamma^{\pm 0.3}$ and $(1-\xgamma)^{\pm 0.3}$ in both
kinematic regions and e) the $y$ distribution by $y^{\pm 0.5}$ and $(1-y)^{\pm
0.5}$ in both kinematic regions.  In DIS, the largest deviation ($9\%$) is due to
the $\xpomeron$ reweighting.  In photoproduction the largest error ($6\%$) arises
from the $\pthat$ reweighting.
\item  The $t$ distribution is varied by factors $e^{\pm 2t/{\rm GeV}^2}$ as
constrained by inclusive measurements~\cite{tslope,kapishin} leading to cross
section errors of $2$ to $3\%$.
\item The estimated number of non-diffractive background events which migrate
into the sample from the unmeasured region $\xpomeron > 0.03$ or $M_Y>5$~GeV is
varied by $\pm 50\%$, leading to a mean cross section uncertainty of $2\%$ in
photoproduction and $3\%$ in DIS.
\item A $7\%$ error arises from uncertainties in the migrations about the $M_Y$
boundary of the measurement.  It is estimated by varying the simulated
efficiencies of the forward detectors FMD and PRT by $\pm 4\%$ and $\pm 25\%$,
respectively, and by variations in the DIFFVM simulation of a) the ratio of
elastic proton to proton dissociation cross sections between $1:2$ and $2:1$, b)
the generated $M_Y$ distribution within $M_Y^{-2.0\pm 0.3}$, c) the $t$
dependence in the proton dissociation simulation by factors $e^{\pm t/{\rm
GeV}^2}$.
\item The loss of diffractive events due to the $\etamax$ cut and the cuts on
the FMD and PRT is corrected using the RAPGAP simulation.  By studying jet
events with an elastically scattered proton (measured in a Roman pot detector)
in the range $\xpom < 0.05$, it is established that the RAPGAP simulation
describes the loss seen in the data within a $10\%$ and $14\%$ statistical
uncertainty for photoproduction and DIS, respectively~\cite{schenk}.  This
uncertainty is used to estimate the uncertainty on the rapidity gap selection
in the present analysis and translates into cross section errors of $1\%$ in both
photoproduction and DIS.
\end{itemize}

The largest errors in photoproduction arise from the uncertainty in the LAr
energy scale and the migrations about the $M_Y$ boundary.  In DIS, the largest
error arises from the $\xpomeron$ reweighting of RAPGAP.  The uncertainties due
to the LAr hadronic energy scale, the energy contribution of tracks, the
luminosity, the FMD noise, the estimated number of background events and the
positron energy in the SPACAL for DIS are correlated between cross section
bins.  Both for the bin-to-bin correlated and the uncorrelated errors all
individual contributions are added in quadrature to obtain the full
uncertainties.

\section{Results}
The measurement results are presented in \figrange{fig_diszpomeron}{fig_scigp}
and are listed in \tabrange{tab_xsdis1}{tab_xsgp2} as bin-averaged differential
hadron level cross sections for a set of kinematic variables which characterise
the scattering process.  The measurements are compared with next-to-leading
order QCD predictions based on the factorisation approach in
\secrange{sec_dis}{sec_xgamdep} and to leading order soft colour neutralisation
models in \secref{sec_sci}.

\subsection{Diffractive dijet production in DIS}
\label{sec_dis}
\begin{figure}[hhh]                    
\centering
\includegraphics[width=0.8\textwidth,keepaspectratio]{%
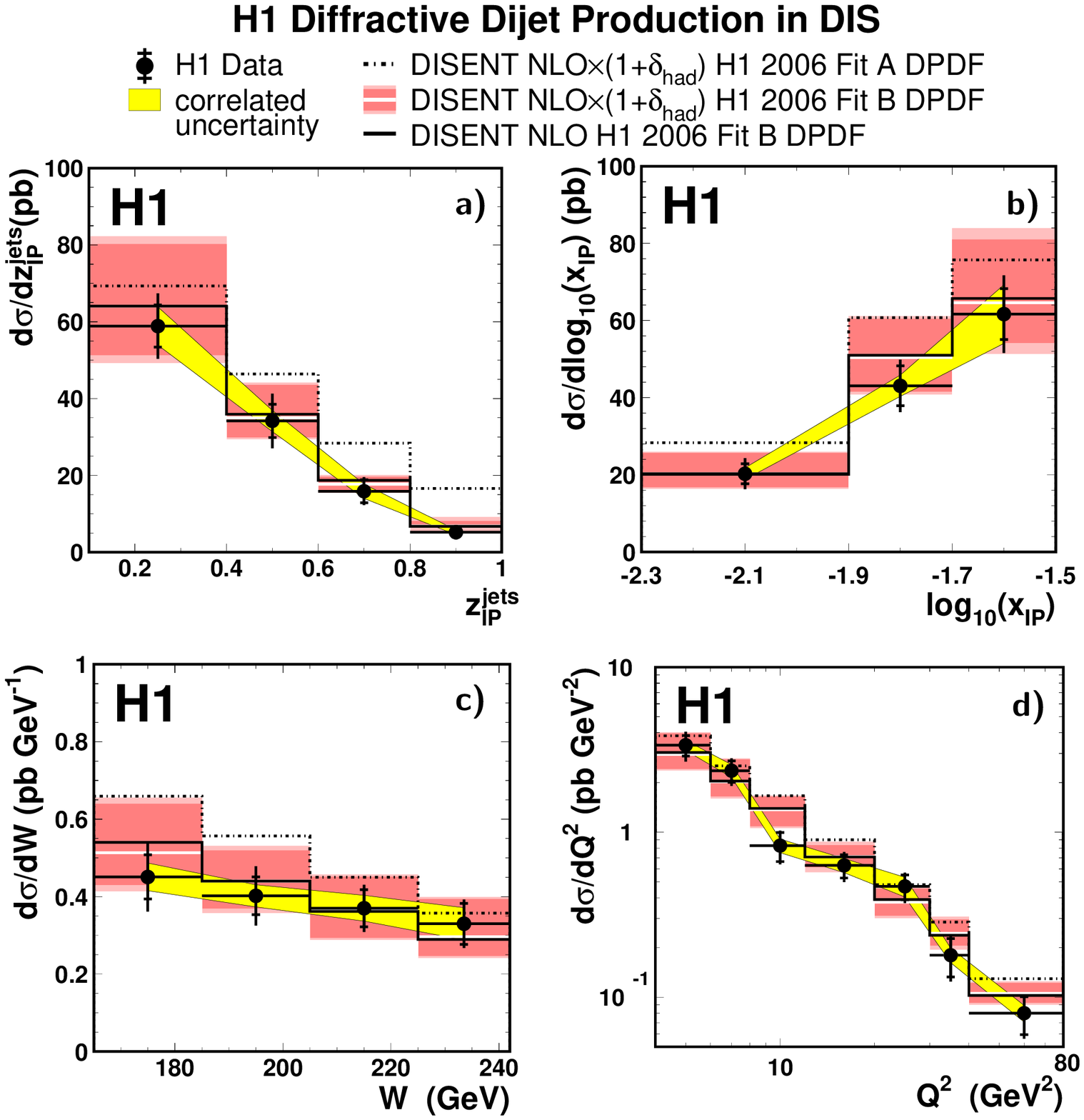}
\caption{Differential cross sections for the diffractive production of two jets
in DIS in the kinematic region specified in \tabref{tab_xsdef} as a function of
a) \zpomeronjets, b) $\logxpomeron$, c) $W$ and d) $Q^2$.  The inner error bars
represent the statistical errors. The outer error bars include the uncorrelated
systematic errors added in quadrature. The shaded band around the data points
indicates an additional systematic uncertainty which is correlated between the
data points.  The predictions based on the QCD program DISENT, using the `H1
2006 Fit A' diffractive parton densities and corrected for hadronisation
effects are shown as the dash-dotted lines.  The predictions based on the `H1
2006 Fit B' DPDFs are shown both with hadronisation corrections (solid white
line) and at the parton level (solid black line). The inner band around the Fit
B predictions indicates the uncertainty resulting from the variation of the
renormalisation scale by factors $0.5$ and $2$ and the full band includes the
uncertainty due to the hadronisation corrections added linearly.}
\label{fig_diszpomeron}
\end{figure}

\begin{figure}[hhh]                    
\centering
\includegraphics[width=0.8\textwidth,keepaspectratio]{%
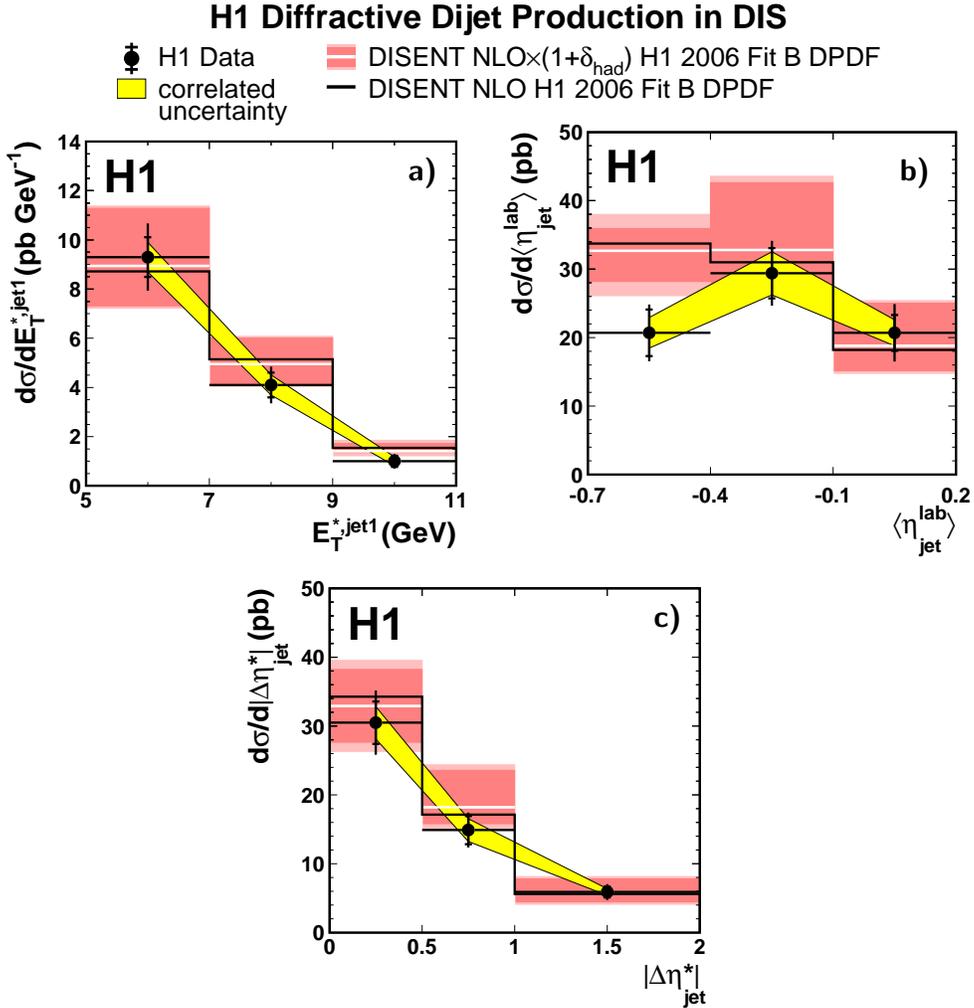}
\caption{Differential cross sections for the diffractive production of two jets
in DIS in the kinematic region specified in \tabref{tab_xsdef} as a function of
the variables a) \ptjetone{}, b) \meanetajetlab{} and c) $\deltaetastar$.  The
DISENT prediction based on the `H1~2006~Fit B' DPDFs  at NLO with (white line)
and without (black line) hadronisation corrections is also shown.  For details
about the errors see the caption of \figref{fig_diszpomeron}.}
\label{fig_disjets}
\end{figure}

In \figsref{fig_diszpomeron}{fig_disjets}, the differential cross sections are
shown as functions of \zpomeronjets, $\logxpomeron$, $W$, $Q^2$, \ptjetone,
\meanetajetlab, and $\deltaetastar$.  The data are compared with NLO QCD
predictions obtained using the DISENT program with the `H1 2006 Fit A' and `H1
2006 Fit B' diffractive parton densities. 

The NLO prediction based on the `H1 2006 Fit A' parton densities (only shown in
\figref{fig_diszpomeron}) overestimates the measured cross section, in
particular at high $\zpomeronjets$.  The NLO prediction based on the `H1 2006
Fit B' parton densities agrees well with the distributions of all variables
within the given errors.  Hence the dijet cross sections distinguish between
the two parton density sets which describe inclusive diffractive DIS similarly
well.  The good description of the differential cross section as a function of
$\logxpomeron$ indicates that the $\xpomeron$ dependence of
$f_{\pom}(x_\pom,t)$ is compatible with the dijet production mechanism within
the shown errors.  The agreement between predicted and measured differential
cross sections as functions of $\ptjetone$ and $\deltaetastar$ suggests that
the NLO QCD matrix element describes the hard scatter correctly within the
uncertainties shown.  The good description of both inclusive diffractive
scattering and diffractive dijet production obtained from the `H1 2006 Fit B'
parton densities supports the validity of QCD hard scattering factorisation in
diffractive DIS.  In the following discussion of diffractive dijet
photoproduction, only the `H1 2006 Fit B' densities are considered.

\subsection{Diffractive photoproduction of dijets}
\label{sec_gp}
\begin{figure}[hhh]                    
\centering
\includegraphics[width=0.8\textwidth,keepaspectratio]{%
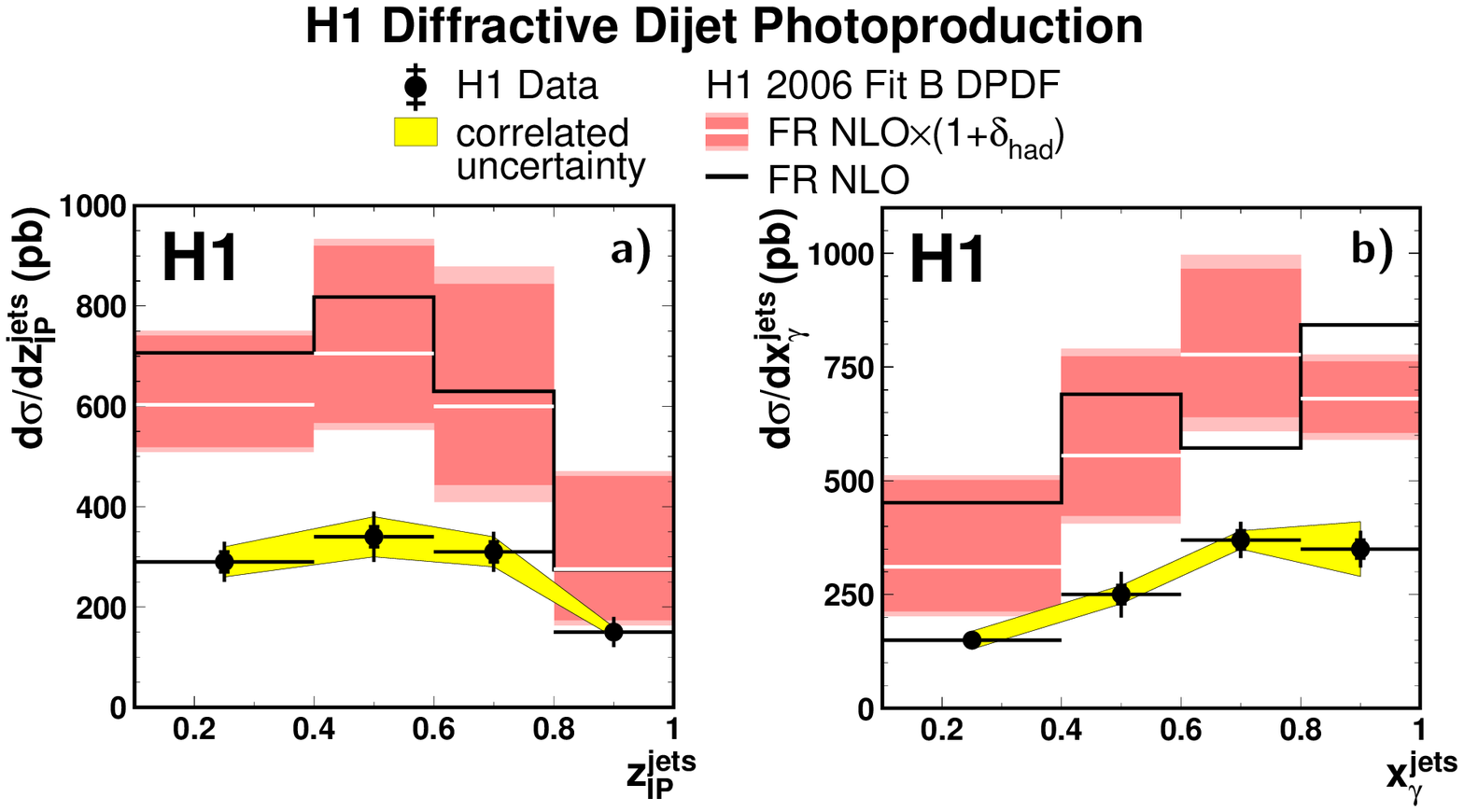}
\caption{Differential cross sections for the diffractive production of two jets
in photoproduction in the kinematic region specified in \tabref{tab_xsdef} as a
function of a) \zpomeronjets{} and b) \xgammajets.  The inner error bars
represent the statistical errors, the outer error bars include the uncorrelated
systematic errors added in quadrature. The shaded band around the data points
indicates an additional systematic uncertainty which is correlated between the
data points.  The NLO QCD predictions based on the Frixione et al. program (FR)
and using the `H1 2006 Fit B' diffractive parton densities are shown with
hadronisation corrections (white line) and at the parton level (black line).
The inner band around the NLO prediction indicates the uncertainty resulting
from simultaneous variations of the renormalisation and factorisation scales by
factors $0.5$ and $2$ and the full band includes the uncertainty due to the
hadronisation corrections added linearly.}
\label{fig_gpzpomeron}
\end{figure}

\begin{figure}[hhh]                    
\centering
\includegraphics[width=0.8\textwidth,keepaspectratio]{%
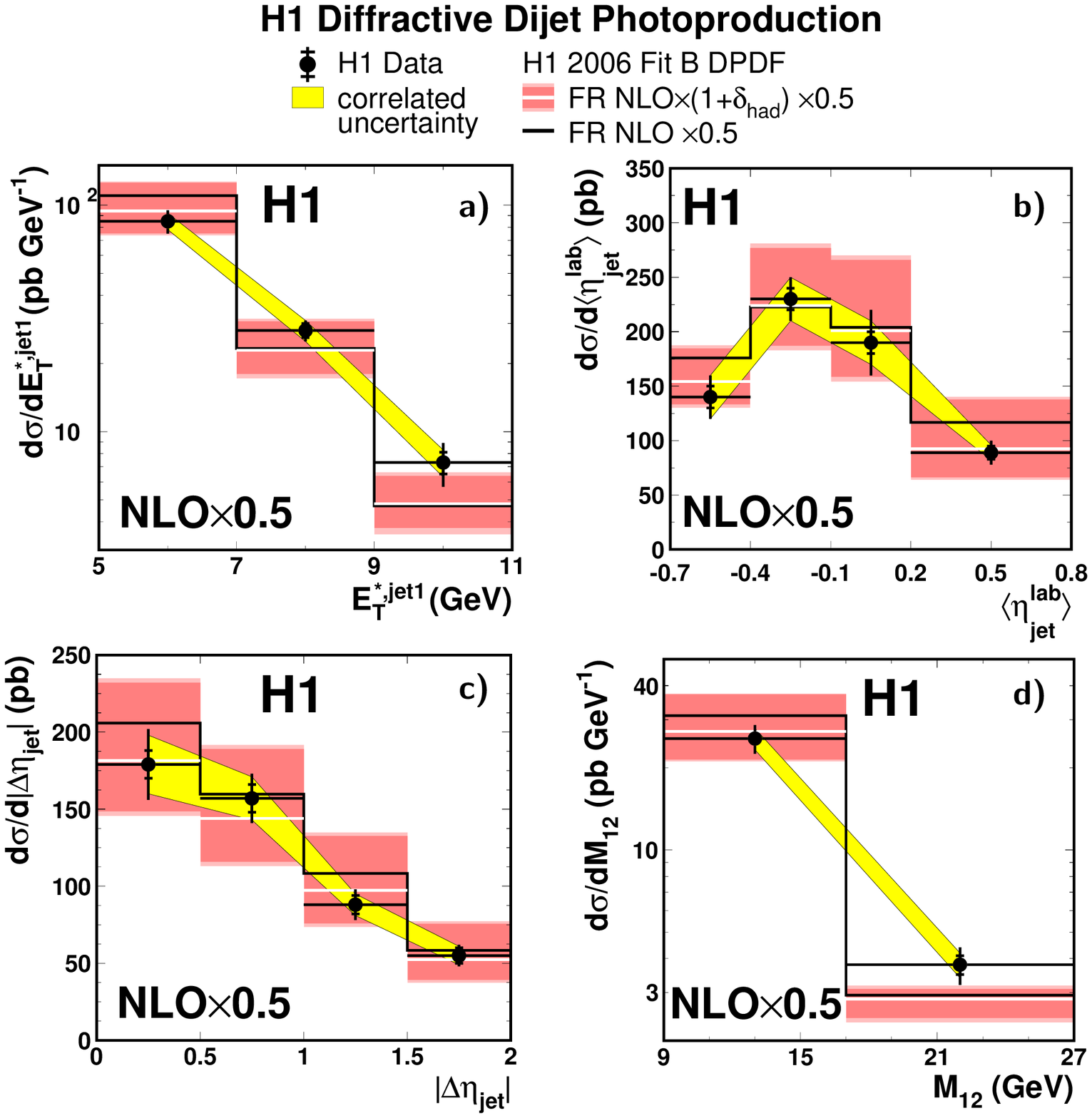}
\caption{Differential cross sections for the diffractive production of two jets
in photoproduction in the kinematic region specified in \tabref{tab_xsdef} as a
function of the jet variables a) \ptjet, b) \meanetajetlab, c) \deltaeta{} and
d) \mjj.  The NLO prediction of the Frixione et al. program interfaced to the
`H1~2006~Fit~B' DPDFs  with and without hadronisation corrections, scaled by an
overall normalisation factor $0.5$ is also shown.  For details about the errors
see the caption of \figref{fig_gpzpomeron}.}
\label{fig_gpjetsscaled}
\end{figure}

Differential cross sections measured for photoproduction are shown
in~\figref{fig_gpzpomeron} as functions of \zpomeronjets{} and $\xgammajets$.
The measurements are compared with NLO predictions obtained with the Frixione
et al. program, interfaced to the `H1~2006~Fit~B' diffractive parton densities.

The NLO prediction overestimates the measured dijet cross section by
a factor of approximately $2$. Diffractive dijet photoproduction thus cannot be described using
the parton densities which lead to a good description of diffractive scattering
in DIS.  QCD hard scattering factorisation is therefore broken in
photoproduction. A more detailed comparison of the cross sections in DIS and
photoproduction is given in the next section.

\subsection{Ratio of dijet cross sections in diffractive photoproduction and DIS}
\label{sec_ratio}
\begin{figure}[hhh]                    
\centering
\includegraphics[width=0.45\textwidth,keepaspectratio]{%
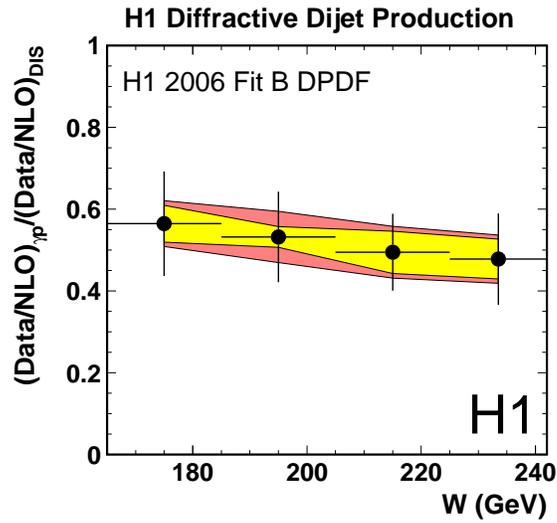}
\caption{Cross section double ratio of data to NLO prediction for
photoproduction and DIS as a function of the photon-proton centre-of-mass
energy $W$.  The error bars indicate uncorrelated experimental uncertainties.
The error bands around the ratio points show systematic uncertainties which are
correlated between the ratio points.  The inner band shows experimental
uncertainties.  The full band shows the quadratic sum of the correlated
experimental uncertainties and NLO QCD uncertainties, estimated from variations
of the factorisation and renormalisation scales.  The nominal QCD scale $E_T$
is varied by the same factors ($0.5$ and $2$) and simultaneously in the same
direction for the DIS and photoproduction calculations.  The two NLO
predictions are based on the same  `H1~2006~Fit~B' diffractive parton densities
and  are corrected for hadronisation effects.} 
\label{fig_ratio}
\end{figure}

A reliable  method to test QCD factorisation is obtained by dividing the ratio
of measured to predicted cross sections in photoproduction by the corresponding
ratio in DIS. In this double ratio many experimental errors and also
theoretical scale errors cancel to a large extent.  The double ratio is shown in
\figref{fig_ratio} as a function of the photon-proton centre-of-mass energy
$W$.  The two NLO calculations are based on the `H1 2006 Fit B' diffractive
parton densities and are corrected for hadronisation. The double ratio is
rather insensitive to the detailed shape of the diffractive gluon density and
the conclusions remain unchanged if the `H1 2006 Fit A' parton densities are
used. 

The double ratio is $\approx 0.5$ throughout the measured $W$ range, indicating
a suppression factor which is independent of the centre-of-mass energy within
the uncertainties.  Integrated over the measured kinematic range the ratio of
data to NLO expectation for photoproduction is a factor $0.5\pm 0.1$ smaller
than the same ratio in DIS where the error includes scale uncertainties. This
confirms that QCD hard scattering factorisation is broken for diffractive dijet
production in photoproduction with respect to the same process in DIS.  The
suppression in photoproduction is much smaller than the suppression in
diffractive dijet production at the Tevatron~\cite{tevjets}.

\subsection{Study of QCD factorisation breaking in photoproduction}
\label{sec_xgamdep}
\begin{figure}[hhh]                    
\centering
\includegraphics[width=0.8\textwidth,keepaspectratio]{%
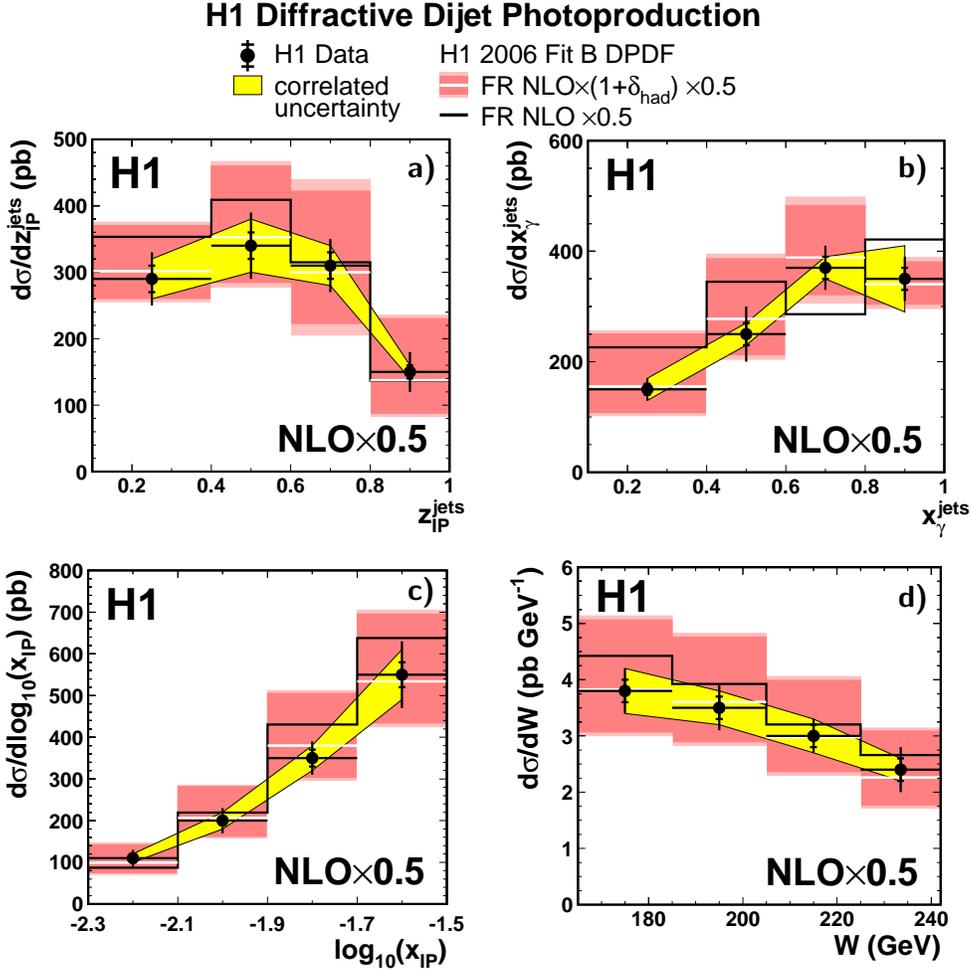}
\caption{Differential cross sections for the diffractive production of two jets
in the photoproduction kinematic region specified in \tabref{tab_xsdef} as a
function of a) \zpomeronjets, b) \xgammajets{}, c) \logxpomeron{} and d) $W$.
The NLO prediction of the Frixione et al. program interfaced to the
`H1~2006~Fit~B' DPDFs  with and without hadronisation corrections, scaled by an
overall normalisation factor $0.5$ is also shown.  For details about the errors
see the caption of \figref{fig_gpzpomeron}. }
\label{fig_gpzpomeronscaled}
\end{figure}

The simple assumption that the suppression factor in photoproduction does not
depend on any kinematic variable is studied by scaling the NLO predictions by
an overall suppression factor $0.5$.  Using such a global factor for both
resolved and direct photon processes leads to a good description of all
measured distributions as shown in
\figsref{fig_gpzpomeronscaled}{fig_gpjetsscaled}.

Whilst a suppression of resolved photoproduction is generally expected, a
suppression of the direct photon contribution is in contradiction to
theoretical expectations~\cite{Collins}. At NLO, the contributions of direct
and resolved photon processes to the dijet cross section cannot be calculated
separately. The following discussion therefore focuses on the dependence of the
suppression on the variable $\xgammajetspar$, reconstructed at the parton level
(PL) from parton jets before hadronisation, which is related to the fraction of
the photon energy entering the jet system.  In events with $\xgammajetspar>0.9$
almost the entire photon energy enters the jet system, whereas for events with
$\xgammajetspar<0.9$ a significant photon remnant system is present which may
lead to secondary interactions and rapidity gap destruction.  A fit of the NLO
prediction to the cross section differential in $\xgammajets$ with two free
normalisation parameters for contributions from $\xgammajetspar<0.9$ and
$\xgammajetspar>0.9$ yields suppression factors of $0.47 \pm 0.16$ and $0.53
\pm 0.14$, respectively.  This result indicates again that the suppression is
independent of $\xgammajetspar$ and that both direct and resolved contributions
have to be suppressed by the same factor. 

\begin{figure}[hhh]                    
\centering
\includegraphics[width=0.8\textwidth,keepaspectratio]{%
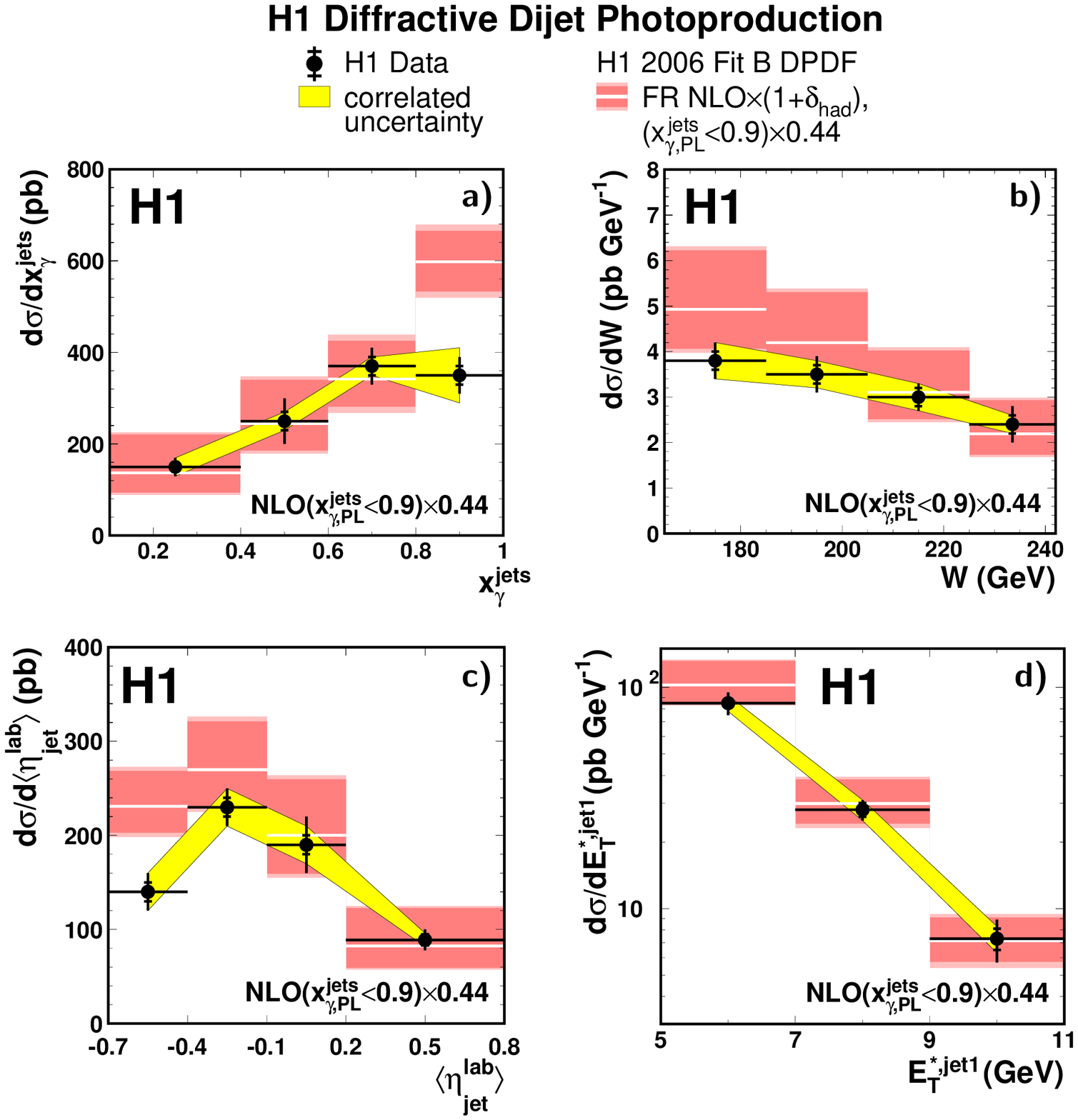}
\caption{Differential cross sections for the diffractive production of two jets
in photoproduction in the kinematic region specified in \tabref{tab_xsdef} as a
function of a) \xgammajets, b) $W$, c) $\meanetajet$ and d) $\ptjet$.  The NLO
prediction of the Frixione et al. program interfaced to the `H1~2006~Fit~B'
DPDFs with hadronisation corrections is also shown. The part of the NLO
calculation for which $\xgammajetspar<0.9$ at the parton level is scaled by
$0.44$.  For details about the errors see the caption of
\figref{fig_gpzpomeron}.}
\label{fig_resscale}
\end{figure}

Finally an investigation is performed of how well the data can be described
under the assumption that the NLO calculation with $\xgammajetspar>0.9$ is not
suppressed.  The best agreement in a $\chi^2$ fit is reached for a suppression
factor $0.44$ for the NLO calculation with $\xgammajetspar<0.9$ and the
resulting distributions are shown for $\xgammajets$, $W$, $\meanetajet$ and
$\ptjet$ in \figref{fig_resscale}. This prediction is incompatible with the
measured cross sections. The assumption that the direct part obeys QCD
factorisation is therefore strongly disfavoured by the present analysis. 

\subsection{Leading order soft colour neutralisation models}
\label{sec_sci}
\begin{figure}[hhh]                    
\centering
\includegraphics[width=0.8\textwidth,keepaspectratio]{%
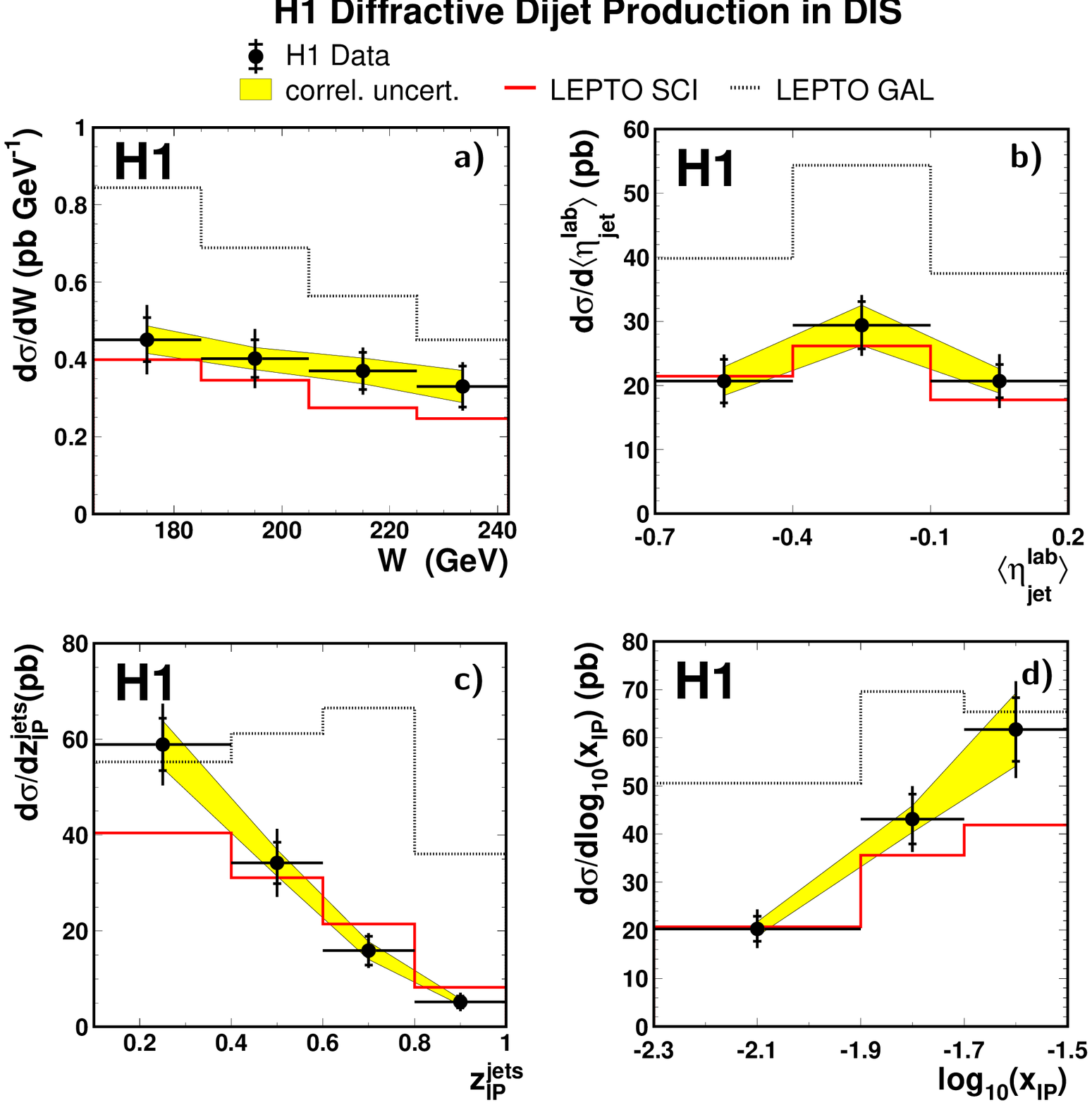}
\caption{Differential cross sections for the diffractive production of two jets
in DIS in the kinematic region specified in \tabref{tab_xsdef} as a function of
the variables a) $W$, b) \meanetajetlab, c) \zpomeronjets{} and d)
\logxpomeron.  Leading order predictions of the soft colour neutralisation
models SCI and GAL as implemented in LEPTO are also shown, based on the CTEQ5L
leading order proton parton densities.  For details about the errors see the
caption of~\figref{fig_diszpomeron}.}
\label{fig_scidis}
\end{figure}

The predictions of the soft colour interaction models SCI and GAL using the
CTEQ5L LO parton densities of the proton are compared with the measurements in
\figref{fig_scidis} in the DIS kinematic region.  The SCI model describes the
dijet cross section reasonably well.  If the GRV '94 HO proton parton
densities~\cite{grv5007} are used the cross sections are underestimated by a
factor of approximately $2$ in agreement with the conclusions drawn in~\cite{disjets}.
The GAL model overestimates the dijet rate by about $65\%$ on average.  It
gives a good description of the shapes of the differential cross sections as
functions of $W$ and $\meanetajetlab$ but not as functions of $\zpomeronjets$
and $\logxpomeron$.

\begin{figure}[hhh]                    
\centering
\includegraphics[width=0.8\textwidth,keepaspectratio]{%
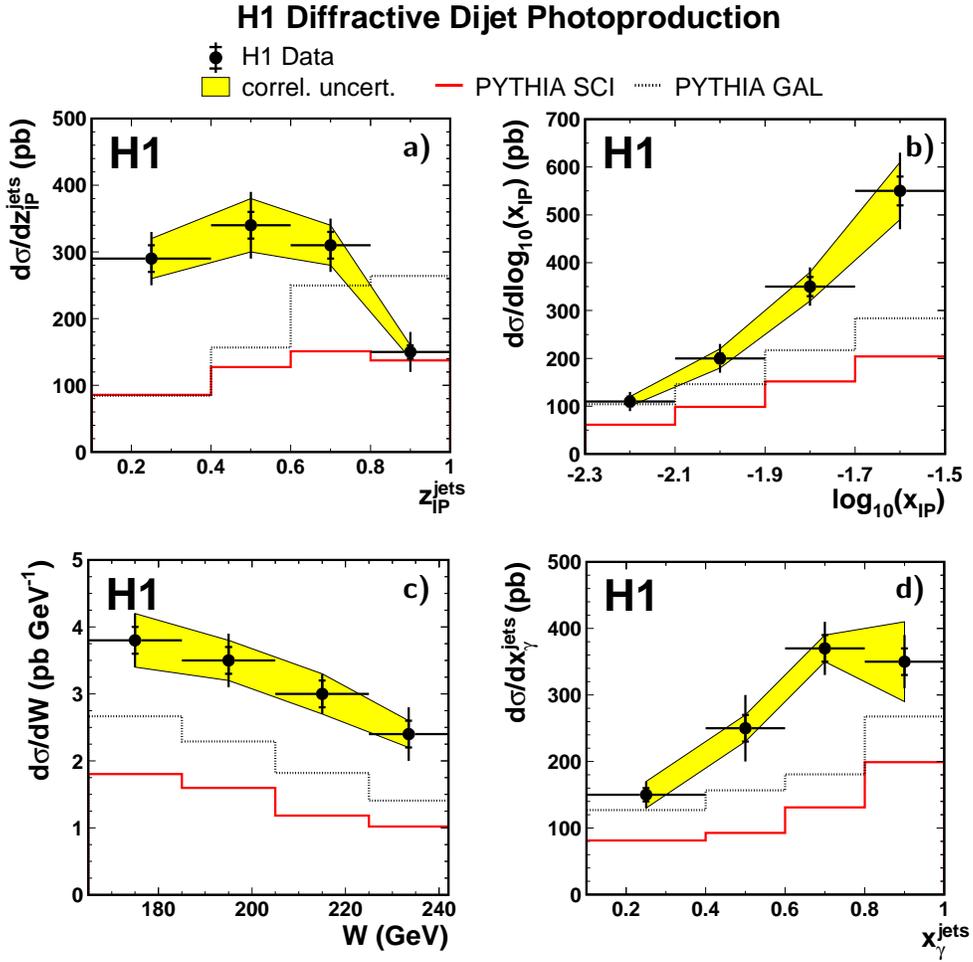}
\caption{Differential cross sections for the diffractive production of two jets
in photoproduction in the kinematic region specified in \tabref{tab_xsdef} as a
function of the variables a) \zpomeronjets, b) \logxpomeron, c) $W$ and d)
\xgammajets.  Leading order predictions of the soft colour neutralisation
models SCI and GAL as implemented in PYTHIA are also shown, based on the CTEQ5L
leading order proton parton densities.}
\label{fig_scigp}
\end{figure}

The predictions for photoproduction are shown in \figref{fig_scigp}. The
normalisation of the cross section is underestimated by factors of
approximately $2.2$
for the SCI model and $1.5$ in the case of the GAL model.  Both models
describe the shapes of the differential cross sections reasonably well for
$\logxpomeron$, $W$ and $\xgammajets$ but fail for $\zpomeronjets$.

In summary, neither of the two models which describe diffractive dijet
production in $p\bar{p}$ collisions is able to describe it in both DIS and
photoproduction.

\section{Summary}
Diffractive dijet production is measured in deep-inelastic scattering and
photoproduction in the same kinematic range \mbox{$165 < W < 242$~GeV},
\mbox{$\xpom < 0.03$,} \mbox{$\etjetone > 5$~GeV} and \mbox{$\etjettwo > 4$
GeV}, with limits on the photon virtuality $4<Q^2<80$~GeV$^2$ for DIS and
$Q^2<0.01$~GeV$^2$ for photoproduction.  The inclusive $k_T$ cluster algorithm
is used in the definition of the jets.

In DIS, diffractive dijet production is well described within the experimental
and theoretical uncertainties by NLO calculations based on diffractive parton
densities determined from QCD fits to inclusive diffractive DIS data.  QCD
factorisation therefore holds within present uncertainties in diffractive DIS.
The dijet measurements clearly favour the `H1 2006 Fit B' over the `H1 2006 Fit
A' parton densities, both of which lead to a good description of inclusive
diffraction.  The gluon densities from the two sets differ mainly for high
fractional momentum.  In this region, the dijet cross section is more sensitive
to the diffractive gluon density than the inclusive scattering cross section.

In photoproduction, NLO calculations based on the `H1 2006 Fit B' parton
densities overestimate the measured cross section.  The ratio of measured cross
section to NLO prediction is a factor $0.5 \pm 0.1$ smaller than the same ratio
in DIS, indicating a clear break-down of QCD factorisation.  A fit to the
photoproduction data yields suppression factors of $0.47\pm 0.16$ for the part
of the NLO calculation for which $\xgammajetspar < 0.9$ and $0.53\pm 0.14$ for
$\xgammajetspar > 0.9$, where $\xgammajetspar$ is the fraction of the photon
momentum entering the hard scatter and is reconstructed at the parton level
from parton jets before hadronisation.  The two factors are compatible with
each other, indicating that the suppression is independent of $\xgammajetspar$.
Direct photon processes contribute primarily at highest values of
$\xgammajetspar$ and the present analysis therefore indicates that they are
suppressed by a similar factor as resolved photon processes.  A suppression of
direct photon processes cannot be explained by models which base the rapidity
gap survival probability on the presence of photon spectator interactions.

The dijet cross sections are also compared with predictions of two soft colour
neutralisation models.  The SCI model which describes diffractive structure
functions at HERA and diffractive dijet production at the Tevatron reproduces
DIS dijet cross sections reasonably well but fails for photoproduction both in
normalisation and in the shape of the differential
cross section in $\zpomeronjets$. The GAL model is incompatible with both data
sets.

\section*{Acknowledgements}
We are grateful to the HERA machine group whose outstanding efforts have made
this experiment possible.  We thank the engineers and technicians for their
work in constructing and maintaining the H1 detector, our funding agencies for
financial support, the DESY technical staff for continual assistance and the
DESY directorate for support and for the hospitality which they extend to the
non DESY members of the collaboration.

\setlength{\tabcolsep}{0.2cm}
\begin{table}                                                                  
\renewcommand{\arraystretch}{1.3}
\begin{center}
{\textbf{Diffractive DIS Dijet Cross Sections}}
\vglue1em
\begin{tabular}{|c|c|c|c|c|c|}
\hline
$\zpomeronjets$  & ${\rm d}\sigma/{\rm d}\zpomeronjets$ (pb) & $\Delta_{\rm stat}$ (pb) & $\Delta_{\rm corr}$ (pb) & $\Delta_{\rm tot}$ (pb) & $1+\delta_{\rm had}$ \\
\hline
$[0.1,0.4)$ & $59$ & $5$ & $5$ & $10$ & $1.00 \pm 0.03$ \\
\hline
$[0.4,0.6)$ & $34$ & $4$ & $3$ & $8$ & $0.97 \pm 0.02$ \\
\hline
$[0.6,0.8)$ & $16$ & $3$ & $2$ & $4$ & $0.95 \pm 0.02$ \\
\hline
$[0.8,1)$ & $5.2$ & $1.2$ & $0.6$ & $2.0$ & $1.0 \pm 0.2$ \\
\hline
\hline
$\logxpomeron$  & ${\rm d}\sigma/{\rm d}\logxpomeron$ (pb) & $\Delta_{\rm stat}$ (pb) & $\Delta_{\rm corr}$ (pb) & $\Delta_{\rm tot}$ (pb) & $1+\delta_{\rm had}$ \\
\hline
$[-2.3,-1.9)$ & $20.3$ & $2.6$ & $1.5$ & $4.3$ & $1.01 \pm 0.02$ \\
\hline
$[-1.9,-1.7)$ & $43$ & $5$ & $3$ & $7$ & $0.99 \pm 0.01$ \\
\hline
$[-1.7,-1.5)$ & $62$ & $7$ & $8$ & $13$ & $0.98 \pm 0.04$ \\
\hline
\hline
$W$ (GeV) & ${\rm d}\sigma/{\rm d}W$  & $\Delta_{\rm stat}$ & $\Delta_{\rm corr}$ & $\Delta_{\rm tot}$ & $1+\delta_{\rm had}$ \\
                                 &        (pb GeV$^{-1}$)    &  (pb GeV$^{-1}$)   &  (pb GeV$^{-1}$) &  (pb GeV$^{-1}$) &  \\
\hline
$[165,185)$ & $0.45$ & $0.06$ & $0.04$ & $0.10$ & $0.95 \pm 0.03$ \\
\hline
$[185,205)$ & $0.40$ & $0.05$ & $0.03$ & $0.08$ & $1.00 \pm 0.03$ \\
\hline
$[205,225)$ & $0.37$ & $0.05$ & $0.03$ & $0.07$ & $1.00 \pm 0.02$ \\
\hline
$[225,242)$ & $0.33$ & $0.05$ & $0.04$ & $0.07$ & $1.03 \pm 0.02$ \\
\hline
\hline
$Q^2$ (GeV$^2$) & ${\rm d}\sigma/{\rm d}Q^2$  & $\Delta_{\rm stat}$ & $\Delta_{\rm corr}$ & $\Delta_{\rm tot}$ & $1+\delta_{\rm had}$ \\
                                 &        (pb GeV$^{-2}$)    &  (pb GeV$^{-2}$)   &  (pb GeV$^{-2}$) &  (pb GeV$^{-2}$) &  \\
\hline
$[4,6)$ & $3.4$ & $0.5$ & $0.2$ & $0.7$ & $0.97 \pm 0.02$ \\
\hline
$[6,8)$ & $2.4$ & $0.4$ & $0.2$ & $0.5$ & $0.99 \pm 0.02$ \\
\hline
$[8,12)$ & $0.83$ & $0.17$ & $0.08$ & $0.21$ & $0.98 \pm 0.02$ \\
\hline
$[12,20)$ & $0.63$ & $0.10$ & $0.06$ & $0.14$ & $1.01 \pm 0.06$ \\
\hline
$[20,30)$ & $0.47$ & $0.08$ & $0.06$ & $0.12$ & $0.96 \pm 0.03$ \\
\hline
$[30,40)$ & $0.18$ & $0.05$ & $0.02$ & $0.06$ & $1.03 \pm 0.05$ \\
\hline
$[40,80)$ & $0.081$ & $0.021$ & $0.009$ & $0.026$ & $1.03 \pm 0.02$ \\
\hline
\end{tabular}
\end{center}
\caption{The hadron level differential cross section of diffractive dijet
production in $ep$ collisions in the DIS kinematic range specified in
\tabref{tab_xsdef}.  The quoted cross section is the average value over the bin
specified in the first column. The quantity $\Delta_{\rm stat}$ is the
statistical uncertainty, $\Delta_{\rm corr}$ the bin-correlated systematic
uncertainty and $\Delta_{\rm tot}$ the total quadratic sum of statistical and
systematic errors including $\Delta_{\rm corr}$. The quantity $1+\delta_{\rm
had}$ is the factor by which the parton level NLO calculation is multiplied to
correct for hadronisation effects.} 
\label{tab_xsdis1}
\end{table}

\setlength{\tabcolsep}{0.2cm}
\begin{table}
\renewcommand{\arraystretch}{1.3}
\begin{center}
{\textbf{Diffractive DIS Dijet Cross Sections}}
\vglue1em
\begin{tabular}{|c|c|c|c|c|c|}
\hline
$\ptjetone$ (GeV) & ${\rm d}\sigma/{\rm d}\ptjetone$  & $\Delta_{\rm stat}$ & $\Delta_{\rm corr}$ & $\Delta_{\rm tot}$ & $1+\delta_{\rm had}$ \\
                                 &        (pb GeV$^{-1}$)    &  (pb GeV$^{-1}$)   &  (pb GeV$^{-1}$) &  (pb GeV$^{-1}$) &  \\
\hline
$[5,7)$ & $9.3$ & $0.8$ & $0.6$ & $1.5$ & $1.03 \pm 0.01$ \\
\hline
$[7,9)$ & $4.1$ & $0.5$ & $0.4$ & $0.9$ & $0.96 \pm 0.01$ \\
\hline
$[9,11)$ & $1.0$ & $0.2$ & $0.2$ & $0.4$ & $0.91 \pm 0.09$ \\
\hline
\hline
$\meanetajetlab$  & ${\rm d}\sigma/{\rm d}\meanetajetlab$ (pb) & $\Delta_{\rm stat}$ (pb) & $\Delta_{\rm corr}$ (pb) & $\Delta_{\rm tot}$ (pb) & $1+\delta_{\rm had}$ \\
\hline
$[-0.7,-0.4)$ & $21$ & $3$ & $2$ & $5$ & $0.97 \pm 0.06$ \\
\hline
$[-0.4,-0.1)$ & $29$ & $4$ & $3$ & $6$ & $1.06 \pm 0.03$ \\
\hline
$[-0.1,0.2)$ & $21$ & $3$ & $2$ & $5$ & $1.03 \pm 0.02$ \\
\hline
\hline
$\deltaetastar$  & ${\rm d}\sigma/{\rm d}\deltaetastar$ (pb) & $\Delta_{\rm stat}$ (pb) & $\Delta_{\rm corr}$ (pb) & $\Delta_{\rm tot}$ (pb) & $1+\delta_{\rm had}$ \\
\hline
$[0,0.5)$ & $30$ & $3$ & $2$ & $5$ & $0.96 \pm 0.04$ \\
\hline
$[0.5,1)$ & $15$ & $2$ & $2$ & $3$ & $1.06 \pm 0.05$ \\
\hline
$[1,2)$ & $5.9$ & $0.8$ & $0.5$ & $1.3$ & $1.01 \pm 0.06$ \\
\hline
\end{tabular}
\end{center}
\caption{The hadron level differential cross section of diffractive dijet
production in $ep$ collisions in the DIS kinematic range specified in
\tabref{tab_xsdef} (continued). For details see the caption of
\tabref{tab_xsdis1}.} 
\label{tab_xsdis2}
\end{table}

\setlength{\tabcolsep}{0.2cm}
\begin{table}                                                                  
\renewcommand{\arraystretch}{1.3}
\begin{center}
{\textbf{Diffractive Photoproduction Dijet Cross Sections}}
\vglue1em
\begin{tabular}{|c|c|c|c|c|c|}
\hline
$\zpomeronjets$  & ${\rm d}\sigma/{\rm d}\zpomeronjets$ (pb) & $\Delta_{\rm stat}$ (pb) & $\Delta_{\rm corr}$ (pb) & $\Delta_{\rm tot}$ (pb) & $1+\delta_{\rm had}$ \\
\hline
$[0.1,0.4)$ & $290$ & $20$ & $30$ & $50$ & $0.85 \pm 0.01$ \\
\hline
$[0.4,0.6)$ & $340$ & $20$ & $40$ & $70$ & $0.86 \pm 0.02$ \\
\hline
$[0.6,0.8)$ & $310$ & $20$ & $30$ & $50$ & $0.95 \pm 0.06$ \\
\hline
$[0.8,1)$ & $150$ & $10$ & $10$ & $30$ & $1.00 \pm 0.04$ \\
\hline
\hline
$\xgammajets$  & ${\rm d}\sigma/{\rm d}\xgammajets$ (pb) & $\Delta_{\rm stat}$ (pb) & $\Delta_{\rm corr}$ (pb) & $\Delta_{\rm tot}$ (pb) & $1+\delta_{\rm had}$ \\
\hline
$[0.1,0.4)$ & $150$ & $10$ & $20$ & $30$ & $0.69 \pm 0.02$ \\
\hline
$[0.4,0.6)$ & $250$ & $20$ & $20$ & $50$ & $0.80 \pm 0.02$ \\
\hline
$[0.6,0.8)$ & $370$ & $20$ & $20$ & $40$ & $1.36 \pm 0.05$ \\
\hline
$[0.8,1)$ & $350$ & $20$ & $60$ & $70$ & $0.81 \pm 0.02$ \\
\hline
\hline
$\logxpomeron$  & ${\rm d}\sigma/{\rm d}\logxpomeron$ (pb) & $\Delta_{\rm stat}$ (pb) & $\Delta_{\rm corr}$ (pb) & $\Delta_{\rm tot}$ (pb) & $1+\delta_{\rm had}$ \\
\hline
$[-2.3,-2.1)$ & $110$ & $10$ & $10$ & $30$ & $1.13 \pm 0.05$ \\
\hline
$[-2.1,-1.9)$ & $200$ & $10$ & $20$ & $30$ & $0.94 \pm 0.02$ \\
\hline
$[-1.9,-1.7)$ & $350$ & $20$ & $30$ & $50$ & $0.88 \pm 0.02$ \\
\hline
$[-1.7,-1.5)$ & $550$ & $30$ & $60$ & $100$ & $0.84 \pm 0.01$ \\
\hline
\hline
$W$ (GeV) & ${\rm d}\sigma/{\rm d}W$  & $\Delta_{\rm stat}$ & $\Delta_{\rm corr}$ & $\Delta_{\rm tot}$ & $1+\delta_{\rm had}$ \\
                                 &        (pb GeV$^{-1}$)    &  (pb GeV$^{-1}$)   &  (pb GeV$^{-1}$) &  (pb GeV$^{-1}$) &  \\
\hline
$[165,185)$ & $3.8$ & $0.2$ & $0.4$ & $0.6$ & $0.87 \pm 0.02$ \\
\hline
$[185,205)$ & $3.5$ & $0.2$ & $0.3$ & $0.5$ & $0.92 \pm 0.02$ \\
\hline
$[205,225)$ & $3.0$ & $0.2$ & $0.3$ & $0.4$ & $0.93 \pm 0.02$ \\
\hline
$[225,242)$ & $2.4$ & $0.2$ & $0.2$ & $0.4$ & $0.85 \pm 0.02$ \\
\hline
\end{tabular}
\end{center}
\caption{The hadron level differential cross section of diffractive dijet
production in $ep$ collisions in the photoproduction kinematic range specified
in \tabref{tab_xsdef}. For details see the caption of \tabref{tab_xsdis1}.}
\label{tab_xsgp1}
\end{table}

\setlength{\tabcolsep}{0.2cm}
\begin{table}                                                                  
\renewcommand{\arraystretch}{1.3}
\begin{center}
{\textbf{Diffractive Photoproduction Dijet Cross Sections}}
\vglue1em
\begin{tabular}{|c|c|c|c|c|c|}
\hline
$\ptjet$ (GeV) & ${\rm d}\sigma/{\rm d}\ptjet$  & $\Delta_{\rm stat}$ & $\Delta_{\rm corr}$ & $\Delta_{\rm tot}$ & $1+\delta_{\rm had}$ \\
                                 &        (pb GeV$^{-1}$)    &  (pb GeV$^{-1}$)   &  (pb GeV$^{-1}$) &  (pb GeV$^{-1}$) &  \\
\hline
$[5,7)$ & $85$ & $3$ & $7$ & $12$ & $0.85 \pm 0.01$ \\
\hline
$[7,9)$ & $28$ & $2$ & $3$ & $4$ & $0.98 \pm 0.03$ \\
\hline
$[9,11)$ & $7.3$ & $0.8$ & $1.0$ & $1.9$ & $1.02 \pm 0.05$ \\
\hline
\hline
$\meanetajetlab$  & ${\rm d}\sigma/{\rm d}\meanetajetlab$ (pb) & $\Delta_{\rm stat}$ (pb) & $\Delta_{\rm corr}$ (pb) & $\Delta_{\rm tot}$ (pb) & $1+\delta_{\rm had}$ \\
\hline
$[-0.7,-0.4)$ & $140$ & $10$ & $20$ & $30$ & $0.88 \pm 0.02$ \\
\hline
$[-0.4,-0.1)$ & $230$ & $10$ & $20$ & $30$ & $1.01 \pm 0.02$ \\
\hline
$[-0.1,0.2)$ & $190$ & $10$ & $20$ & $30$ & $0.99 \pm 0.02$ \\
\hline
$[0.2,0.8)$ & $89$ & $6$ & $7$ & $13$ & $0.79 \pm 0.02$ \\
\hline
\hline
$\deltaeta$  & ${\rm d}\sigma/{\rm d}\deltaeta$ (pb) & $\Delta_{\rm stat}$ (pb) & $\Delta_{\rm corr}$ (pb) & $\Delta_{\rm tot}$ (pb) & $1+\delta_{\rm had}$ \\
\hline
$[0,0.5)$ & $179$ & $9$ & $19$ & $29$ & $0.88 \pm 0.01$ \\
\hline
$[0.5,1)$ & $157$ & $9$ & $14$ & $21$ & $0.90 \pm 0.02$ \\
\hline
$[1,1.5)$ & $88$ & $6$ & $7$ & $13$ & $0.90 \pm 0.02$ \\
\hline
$[1.5,2)$ & $55$ & $5$ & $6$ & $9$ & $0.90 \pm 0.03$ \\
\hline
\hline
$M_{12}$ (GeV) & ${\rm d}\sigma/{\rm d}M_{12}$  & $\Delta_{\rm stat}$ & $\Delta_{\rm corr}$ & $\Delta_{\rm tot}$ & $1+\delta_{\rm had}$ \\
                                 &        (pb GeV$^{-1}$)    &  (pb GeV$^{-1}$)   &  (pb GeV$^{-1}$) &  (pb GeV$^{-1}$) &  \\
\hline
$[9,17)$ & $25.6$ & $0.9$ & $2.3$ & $3.9$ & $0.88 \pm 0.01$ \\
\hline
$[17,27)$ & $3.8$ & $0.3$ & $0.4$ & $0.7$ & $0.97 \pm 0.03$ \\
\hline
\end{tabular}
\end{center}
\caption{The hadron level differential cross section of diffractive dijet
production in $ep$ collisions in the photoproduction kinematic range specified
in \tabref{tab_xsdef} (continued). For details see the caption of
\tabref{tab_xsdis1}.} 
\label{tab_xsgp2}
\end{table}
\end{document}